\documentclass[11pt]{article}
\usepackage[utf8]{inputenc}

\usepackage{graphicx}
\usepackage{amsmath,amssymb}
\usepackage{latexsym}
\usepackage{subfigure}
\usepackage{wrapfig}
\usepackage{color}
\usepackage{float}
\usepackage{hyperref}
\usepackage{comment}
\usepackage{soul}
\usepackage{authblk}
\usepackage{tablefootnote}
\usepackage{multirow}
\usepackage{pdfpages}

\newcommand{\beq}{\begin{equation}}
\newcommand{\eeq}{\end{equation}}
\newcommand{\beqs}{\begin{eqnarray}}
\newcommand{\eeqs}{\end{eqnarray}}

\usepackage[left]{lineno}
\usepackage[backend=biber,style=nature]{biblatex}
\title{Designing shape-memory-like microstructures in intercalation materials}
\author{Delin Zhang}
\author{Ananya Renuka Balakrishna\thanks{Corresponding author: renukaba@usc.edu}}
\affil{\small{Aerospace and Mechanical Engineering, University of Southern California, Los Angeles, CA 90089, USA}}
\date{}
\begin{document}

\maketitle
\begin{abstract}
During the reversible insertion of ions, lattices in intercalation materials undergo structural transformations. These lattice transformations generate misfit strains and volume changes that, in turn, contribute to the structural decay of intercalation materials and limit their reversible cycling. In this paper, we draw on insights from shape-memory alloys, another class of phase transformation materials, that also undergo large lattice transformations but do so with negligible macroscopic volume changes and internal stresses. We develop a theoretical framework to predict structural transformations in intercalation compounds and establish crystallographic design rules necessary for forming shape-memory-like microstructures in intercalation materials. We use our framework to systematically screen open-source structural databases comprising $n>5,000$ pairs of intercalation compounds. We identify candidate compounds, such as Li$_x$Mn$_2$O$_4$ (Spinel), Li$_x$Ti$_2$(PO$_4$)$_3$ (NASICON), that approximately satisfy the crystallographic design rules and can be precisely doped to form shape-memory-like microstructures. Throughout, we compare our analytical results with experimental measurements of intercalation compounds. We find a direct correlation between structural transformations, microstructures, and increased capacity retention in these materials. These results, more generally, show that crystallographic designing of intercalation materials could be a novel route to discovering compounds that do not decay with continuous usage. 
\end{abstract}

\section*{Introduction}
\begin{figure}[ht]
    \centering
    \includegraphics[width=\textwidth]{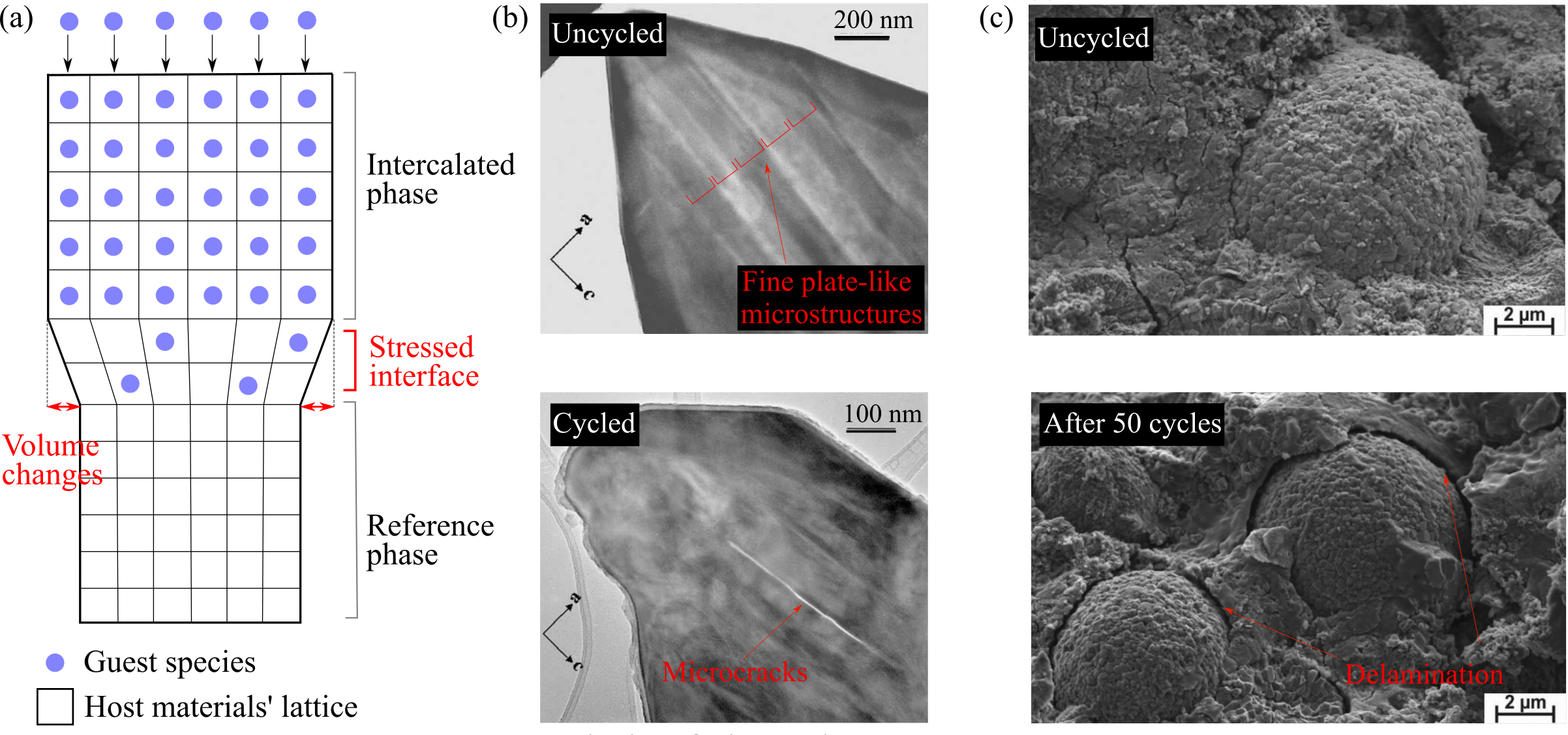}
    \caption{(a) Schematic illustration of intercalation during which guest species are reversibly inserted into a material's lattice structure. This insertion induces structural transformation of the material resulting in stressed interfaces and volume changes. On repeated intercalation, (b) the material's stressed interfaces nucleate microcracks \cite{chen2006electron} (\textcolor{black}{Reprinted with permission from Electrochemical Society and Institute of Electrical and Electronics Engineers}) and (c) its volume changes lead to delamination \cite{koerver2017capacity} (\textcolor{black}{Reprinted with permission from American Chemical Society}). These chemo-mechanical degradations eventually lead to a decay in material performance.}
    \label{Fig1}
\end{figure}

Intercalation is the reversible insertion of guest species (e.g., molecules, atoms or ions) into a material's lattice structure, see Fig.~\ref{Fig1}(a). This reversible insertion makes intercalation materials well-suited for sustainable energy storage, such as graphite in hydrogen storage, electrodes in lithium batteries, and chalcogenides in electrochromic applications \cite{dresselhaus1981intercalation, whittingham1978chemistry, padhi1997phospho, liu2020spontaneous, zhou2014two, lim2016intercalation, nadkarni2019modeling}. This intercalation, however, is typically accompanied by an abrupt structural transformation of the material that shortens its lifespan.

At the microscopic scale, this intercalation-induced transformation leads to a misfit between neighboring lattices which, in turn, leads to a stressed interface. At the macroscopic scale, this transformation induces volume changes of the intercalation material \cite{lewis2019chemo} and leads to non-uniform intercalation behavior \cite{zhang2020stress}. These unwanted features can nucleate microcracks \cite{chen2006electron}, see Fig.~\ref{Fig1}(b); result in mechanically damaged surfaces (delamination), see Fig.~\ref{Fig1}(c) \cite{koerver2017capacity,bucci2018mechanical}; and, in extreme cases, lead to the amorphization of the intercalation material \cite{xiang2017accommodating, zhang2021film}. The structural transformation and its accompanying coherency stress thus contribute to the decay of intercalation materials, which, as a result, need to be replaced \cite{balakrishna2022crystallographic}. However, common applications such as lithium batteries require these materials to survive thousands of intercalation cycles.

\begin{figure}[ht]
    \centering
    \includegraphics[width=\textwidth]{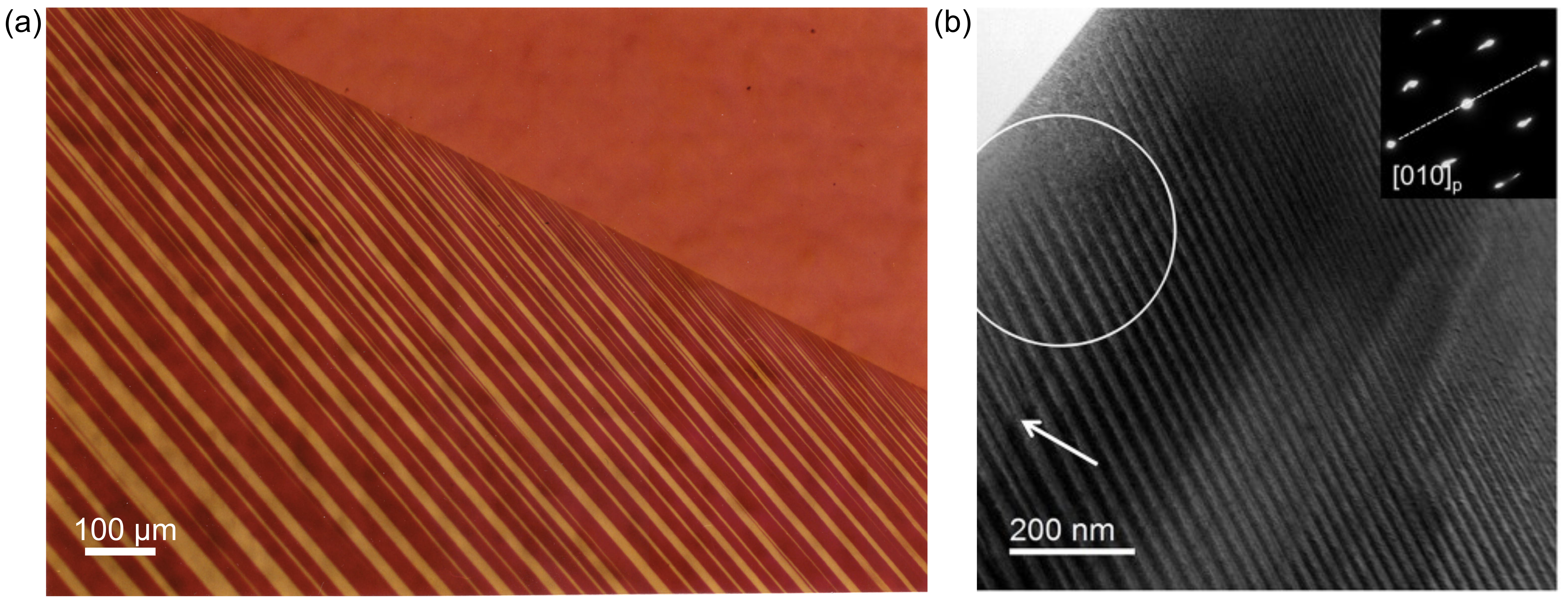}
    \caption{Phase transformation microstructures in intercalation materials bear a striking resemblance to microstructures in shape memory alloys. (a) An austenite-martensite interface showing finely twinned domains in the $\mathrm{Cu}$-14.0\%$\mathrm{Al}$-3.9\%$\mathrm{Ni}$ shape-memory alloy. Image courtesy of Chu and James \cite{chu1995analysis}. (b) A bright field image of a partially transformed Li$_x\mathrm{Mn_2O_4}$ showing finely twinned domains. Reproduced with permission from American Chemical Society \cite{erichsen2020tracking}.}
    \label{Fig2}
\end{figure}

In shape memory alloys (SMAs), another class of phase transformation materials, the structural changes of lattices are also accompanied by large strains. Despite the large transformation strains, SMAs form characteristic microstructures with small coherency stresses (austenite-martensite interface), negligible volume changes (self-accommodating), and low fatigue ($\lambda_2 = 1$). In these materials, lattices of different orientations rotate and/or shear to fit with each other resulting in the formation of twin boundaries, see Fig.~\ref{Fig2}(a) \cite{chu1995analysis}. These finely twinned microstructures minimize coherency stresses at the phase boundary, and, for specific geometric conditions, adapt to the macroscopic material shape. These microstructures, if stabilized in intercalation electrodes, could mitigate the chemo-mechanical challenges plaguing solid-state battery materials.

Intercalation materials and SMAs share several similarities. First, during phase transformation, lattices in both materials undergo reversible structural changes at a critical point. For example, intercalation electrodes transform between the intercalated (lithiated) and reference (delithiated) phases at critical intercalant (guest-species such as Li) concentrations, while SMAs transform between the austenite and martensite phases at the Curie temperature, see Fig.~\ref{Fig2}. Second, during phase transformations, atoms of the host intercalation material and atoms of the SMA undergo cooperative and homogeneous displacements. For example, an intercalation material's lattices undergo structural changes during Li-diffusion, which is analogous to the SMA's lattices undergoing structural changes during thermal diffusion \cite{rudraraju2016mechanochemical}. 

These structural changes in the host intercalation material are supported by, in-situ, X-ray diffraction experiments that show abrupt lattice transformations. Recently, researchers imaged a nanotwinned microstructure in a Spinel cathode ($\mathrm{Li_\mathit{x}Mn_2O_4}$) that forms during electrochemical lithiation (discharging), see Fig.~\ref{Fig2}(b) \cite{erichsen2020tracking}. These nanotwinned microstructures resemble the austenite-martensite interface in SMAs, and are formed to relieve coherency stresses at the phase boundary. In another study, researchers showed that ordered microstructures with plate-like features form in $\mathrm{LiFePO}_4$ electrodes to lower stresses in the particle geometry \cite{chen2006electron}. While these elementary features of twinned microstructures have been observed in some intercalation materials, shape-memory-like microstructures with self-accommodating or low-fatigue characteristics have not.

The self-accommodating and low-fatigue (or $\lambda_2 = 1$) microstructures have important advantages that could address the chemo-mechanical challenges in intercalation materials. The self-accommodation microstructure, adapts to the original shape of the material without any macroscopic change in volume, despite significant structural changes at the atomic scale \cite{bhattacharya1992self}. The $\lambda_2 = 1$ microstructure corresponds to an exactly compatible and stress-free phase boundary that can move back and forth in the material, reversibly, several thousand times, and yet induce ultra-low fatigue \cite{chen2013study, james2005way}. These microstructures are observed in shape-memory alloys and other functional materials and have contributed to a phenomenal improvement in their performance \cite{chluba2015ultralow}.

The self-accommodating or $\lambda_2 = 1$ microstructures form when the material's lattice parameters satisfy very specific geometric relationships \cite{ball1989fine}. We hypothesize that although some intercalation materials \textit{approximately} satisfy these lattice parameter conditions, the majority of the intercalation compounds do not satisfy the crystallographic geometric constraints that are necessary to form the special microstructures with chemo-mechanical advantages. Additionally, we hypothesize that intercalation materials that approximately satisfy the geometric constraints show improved material performance (e.g., reversible cycling and capacity retention). In this paper, we test these hypotheses by conducting a systematic search of the structural data of intercalation materials and analyzing microstructures that form in these materials during phase transformation.

The central aim of this work is to quantify structural transformations in commonly used intercalation materials and to establish crystallographic design rules necessary to reduce coherency stresses and volume changes. \textcolor{black}{In the Methods section}, we outline our theoretical framework to quantify structural transformation pathways and establish the crystallographic design rules necessary to form shape-memory-like microstructures in intercalation materials. \textcolor{black}{In the Results section}, we use our newly developed framework for two studies: In Study 1, we analyze the structural transformations in commonly used crystal structures (e.g., Layered, Spinel, Olivine) and identify families of intercalation compounds that are capable of forming shape-memory-like microstructures. In Study 2, we apply our algorithms to the Materials Project database to analyze whether any known intercalation compound can form shape-memory-like microstructures during phase transformation. Our results show that none of the existing intercalation compounds exactly satisfy the geometric constraint for self-accommodating or highly reversible microstructures; however, intercalation compound groups, such as the Spinels, Tavorites, Phosphates, and NASICON, satisfy a few of the fundamental design principles. These compounds can be systematically doped either using first-principles calculations or site-selective synthesis to precisely satisfy lattice parameter relationships \cite{urban2016computational, kang2006factors, parker2022alloy, martinolich2020controlling}. Throughout, we compare our analytical results with experimental observations and show a direct correlation between structural transformations, microstructural patterns, and material performance. Our results, more broadly, establish a theoretical framework that enables the discovery of novel intercalation compounds with reduced chemo-mechanical challenges.


\section*{Results}\label{Results}
\begin{figure}[ht]
    \centering
    \includegraphics[width=\textwidth]{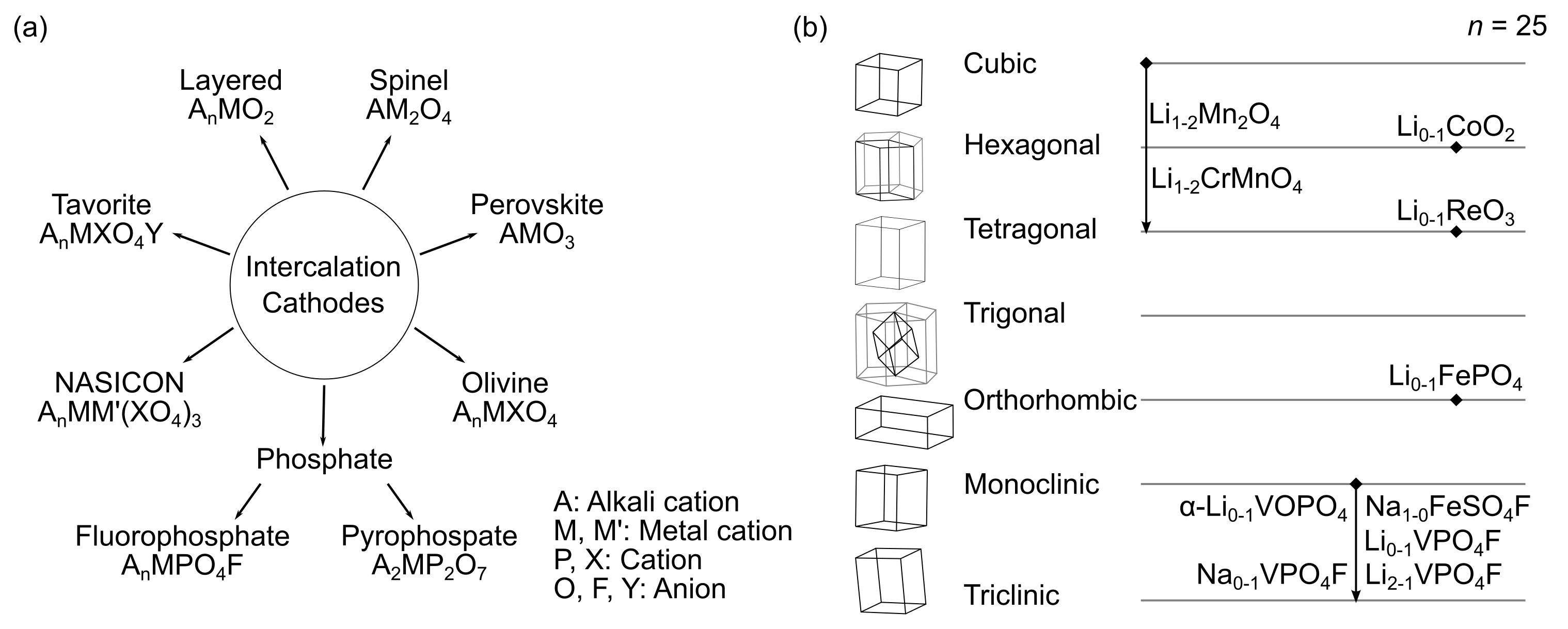}
    \caption{(a) Crystal structures of commonly used intercalation cathodes in lithium batteries. We extract structural data of intercalation cathodes ($n=25$) from these families in Study 1. (b) Our analysis shows that only \textcolor{black}{7} out of 25 intercalation compounds show a change in lattice symmetry during phase transformation.}
    \label{Fig5}
\end{figure}

In Study 1, we apply our theoretical framework to commonly used cathode compounds (see Fig.~\ref{Fig5}(a)), and compare our analysis with microstructural measurements of $\mathrm{Li_\mathit{x}Mn_2O_4}$ and X-ray diffraction measurements for $n=25$ intercalation compounds. In Study 2, we apply our framework to a larger structural database of intercalation compounds (i.e., Materials Project database comprising structural data for over 5,158 pairs of reference/intercalated compounds).\footnote{Intercalation compounds such as $\mathrm{Li_\mathit{x}V_2O_5}$ undergo multiple phase transformations as a function of Li-content. In these cases, we categorize the end products across each phase transformation stage as a distinct pair of intercalation compounds.} We systematically search and analyze this database to determine whether any known intercalation compound satisfies the crystallographic design principles identified in Table~\ref{Table1}. Our findings show that Spinels, Tavorites, Phosphates, and NASICON, are a group of intercalation compounds that approximately satisfy the fundamental design principles necessary to form crystallographic microstructures. We identify specific candidate compounds that approximately satisfy the geometric conditions for forming $\lambda_2 = 1$ and/or self-accommodation microstructures, and could be systematically doped to precisely satisfy the geometric constraints described in Table~\ref{Table1}. Overall, our analysis shows a direct correlation between structural transformations of unit cells, microstructural patterns, and intercalation material performance.

\subsection*{Study 1}
We compute the stretch tensors $\mathbf{U}$ for $n = 25$ representative intercalation compounds, see Supplementary Table~7. These intercalation compounds are the commonly used cathodes from the Layered, Olivine, NASICON, or Spinel families, which have stable structures and large interstitial spaces for reversible ion insertion, see Fig.~\ref{Fig5}(a) \cite{islam2014lithium}. We analyze the lattice geometries of $n=25$ compounds from the X-ray diffraction data and compute their corresponding structural transformation pathways. We find that \textcolor{black}{$\sim28\%$} of the compounds undergo symmetry-lowering structural transformations during charge/discharge cycles, see Fig.~\ref{Fig5} and Supplementary Table~7. This lattice-symmetry lowering transformation is necessary to form twin interfaces and crystallographic microstructures described in Table~\ref{Table1}. Furthermore, we identify that intercalation compounds with Spinel and Pyrophosphate structures approximately satisfy the crystallographic design principles (i.e. form twins, austenite/martensite microstructures) and with precise lattice-geometries can form stress-free phase transformation microstructures. Below we present our microstructural analysis using Li$_x$Mn$_2$O$_4$ as a representative compound. The analyses of other intercalation compounds are presented in the \textcolor{black}{Supplementary Information}.

\subsubsection*{Twin interface}
\textcolor{black}{Li$_\mathit{x}$Mn$_2$O$_4$ (LMO)} is a Spinel compound that undergoes a first-order phase transformation between LiMn$_2$O$_4$ (reference phase) and Li$_2$Mn$_2$O$_4$ (intercalated phase). Using our algorithm, we determine that the cubic-to-tetragonal structural transformation of LMO minimizes the distance function $\left\|\mathbf{U}^{-2}-\mathbf{I}\right\|^2$. This cubic (LiMn$_2$O$_4$) to tetragonal (Li$_2$Mn$_2$O$_4$) transformation generates three variants and their corresponding stretch tensors are listed in Table~\ref{Table2}.\footnote{Variants are lattices of different orientations belonging to the same phase of the material. These variants are related to each other via a symmetry operation.} This transformation pathway, predicted by our algorithm, is consistent with the experimental measurements \cite{erichsen2020tracking}, and the lattice correspondence matrices determined by our code are in agreement with those in the International Tables for Crystallography \cite{brock2016international}.

\begin{figure}[ht]
    \centering
    \includegraphics[width=0.9\textwidth]{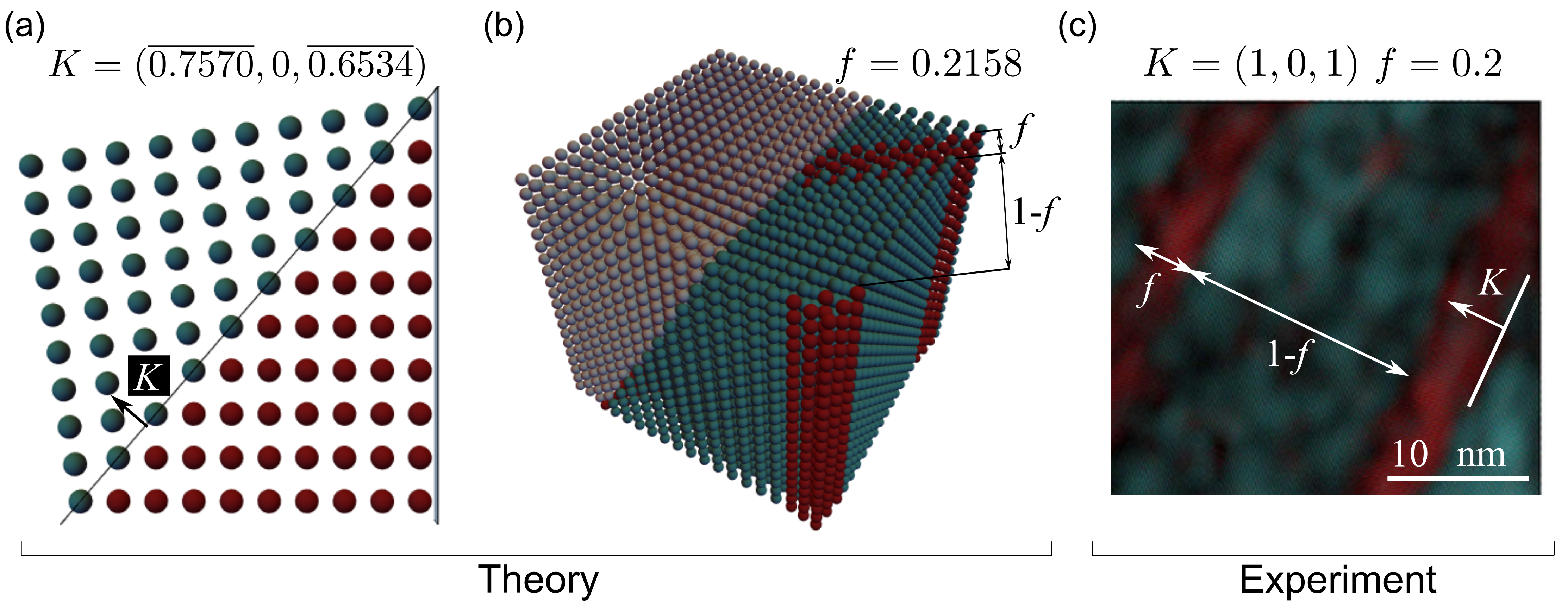}
    \caption{Geometric construction of twin microstructures in $\mathrm{Li_2Mn_2O_4}$. (a) Using the analytically derived twin solutions for $\mathrm{Li_2Mn_2O_4}$ in Table~\ref{Table2}, we geometrically construct a twin interface using variants $\mathbf{U}_2$ and $\mathbf{U}_3$. A cross-sectional view of the twin interface shows a twin-plane $K=(\overline{0.7570},\ 0,\ \overline{0.6534})$. (b) A 3D construction of the austenite/martensite interface in $\mathrm{Li_2Mn_2O_4}$. Here, the cubic-reference phase ($\mathrm{LiMn_2O_4}$) forms a coherent interface with finely twinned tetragonal-intercalated phase ($\mathrm{Li_2Mn_2O_4}$). We predict the volume fraction of the twinned mixture to be $f = 0.2158$. (c) Our geometric construction closely matches the previously imaged (Bright field) LMO sample showing (101) twining plane in Cartesian coordinates and volume fraction $f = 0.2$. Reproduced with permission from American Chemical Society \cite{erichsen2020tracking}.}
    \label{Fig6}
\end{figure}

Any two variants of the tetragonal-intercalated phase satisfy the compatibility condition to form a twin interface, see Table~\ref{Table1}. Table~\ref{Table2} lists the twin solutions, namely vectors $\mathbf{a}$, $\hat{\mathbf{n}}$ and rotation tensors $\mathbf{Q}$, for $\mathbf{U}_2$ and $\mathbf{U}_3$ variants. We use these solutions to geometrically construct the orientation of a twin interface in Fig.~\ref{Fig6}(a). The computed twin plane $K=(\overline{0.7570},\ 0,\ \overline{0.6534})$ closely matches the experimental measurement (1, 0, 1) in Cartesian coordinates, see Fig.~\ref{Fig6}(a-c).

\begin{table}[ht]
    \small
    \centering
    \renewcommand\arraystretch{1.5}
    \addtolength{\leftskip} {-2cm}
    \addtolength{\rightskip}{-2cm}
    \begin{tabular}{lll}
    \hline
    Stretch tensor& Twin solution & A/M solution\\
    \hline
    \\
    $\mathbf{U}_1 =\begin{bmatrix} 0.9690&0&0\\0&1.1226&0\\0&0&0.9690\end{bmatrix}$& &\\[6ex]
    \multirow{3}{15em}{$\mathbf{U}_2 =\begin{bmatrix} 1.1226&0&0\\0&0.9690&0\\0&0&0.9690\end{bmatrix}$} & $\mathbf{a}=[\overline{0.2002},\ 0,\ 0.2319]$, $k=-1$ & $f=0.2158$, $k=1$\\
    &$\mathbf{\hat{n}}=(\overline{0.7071},\ 0,\ \overline{0.7071})$&$\mathbf{b}=[0.0996, 0.0646, \overline{0.0043}]$\\
    &$K=(\overline{0.7570},\ 0,\ \overline{0.6534})$&$\mathbf{\hat{m}}=(0.8653, -0.4998, -0.0376)$\\[1ex]
    
    $\mathbf{U}_3 = \begin{bmatrix} 0.9690&0&0\\0&0.9690&0\\0&0&1.1226\end{bmatrix}$ & $\mathbf{Q}= \begin{bmatrix} 0.9893&0&0.1461\\0&1&0\\-0.1461&0&0.9893\end{bmatrix}$ & $\mathbf{Q'}= \begin{bmatrix} 0.9982&-0.0514&0.0316\\0.0515&0.9987&-0.0006\\-0.0315&0.0022&0.9995\end{bmatrix}$\\[5ex]
    \\
    \hline
    \end{tabular}
    \caption{Microstructural solutions for intercalation cathode Li$_2$Mn$_2$O$_4$. The stretch tensors $\mathbf{U}_1, \mathbf{U}_2, \mathbf{U}_3$ are computed from lattice geometry measurements reported in Ref.~\cite{erichsen2020tracking}. Using variants $\mathbf{U}_2, \mathbf{U}_3$, we construct the solutions to twin interfaces and the austenite/martensite microstructure shown in Fig.~\ref{Fig6}.}
    \label{Table2}
\end{table}

\subsubsection*{Austenite-Martensite microstructure}
The structural transformation of Li$_x$Mn$_2$O$_4$ generates coherency stresses at the phase boundary between the cubic LiMn$_2$O$_4$ phase and the tetragonal Li$_2$Mn$_2$O$_4$ phase. To minimize these coherency stresses, the tetragonal variants form a finely twinned mixture that fits compatibly with the cubic phase and results in the characteristic austenite-martensite microstructure, see Fig.~\ref{Fig6}(c). 

We use the stretch tensors $\mathbf{U}_2$ and $\mathbf{U}_3$ of Li$_x$Mn$_2$O$_4$, computed in Table~\ref{Table2}, to solve for the austenite-martensite microstructure. The stretch tensors satisfy the compatibility condition in Table~\ref{Table1} for a volume fraction $f = 0.2158$.  Table~\ref{Table2} lists a solution of the austenite-martensite interface, namely vectors $\mathbf{b}$, $\hat{\mathbf{m}}$, and the rotation tensors $\mathbf{Q'}$, for corresponding volume fractions $f$.\footnote{\textcolor{black}{Following similar steps, other solutions of the austenite-martensite interface can be obtained between other tetragonal variants, see Supplementary Table~5.}} Our analytical prediction of the volume fraction for the austenite-martensite microstructure $f = 0.2158$ is consistent with the experimental measurements of $f = 0.2$ by Erichsen et al., \cite{erichsen2020tracking}, see Fig.~\ref{Fig6}(a) and (c). Similarly, solutions to the austenite-martensite interface can be computed for other intercalation compounds that undergo symmetry-lowering transformation.
\subsubsection*{Special microstructures}
\begin{figure}[ht]
    \centering
    \includegraphics[width=\textwidth]{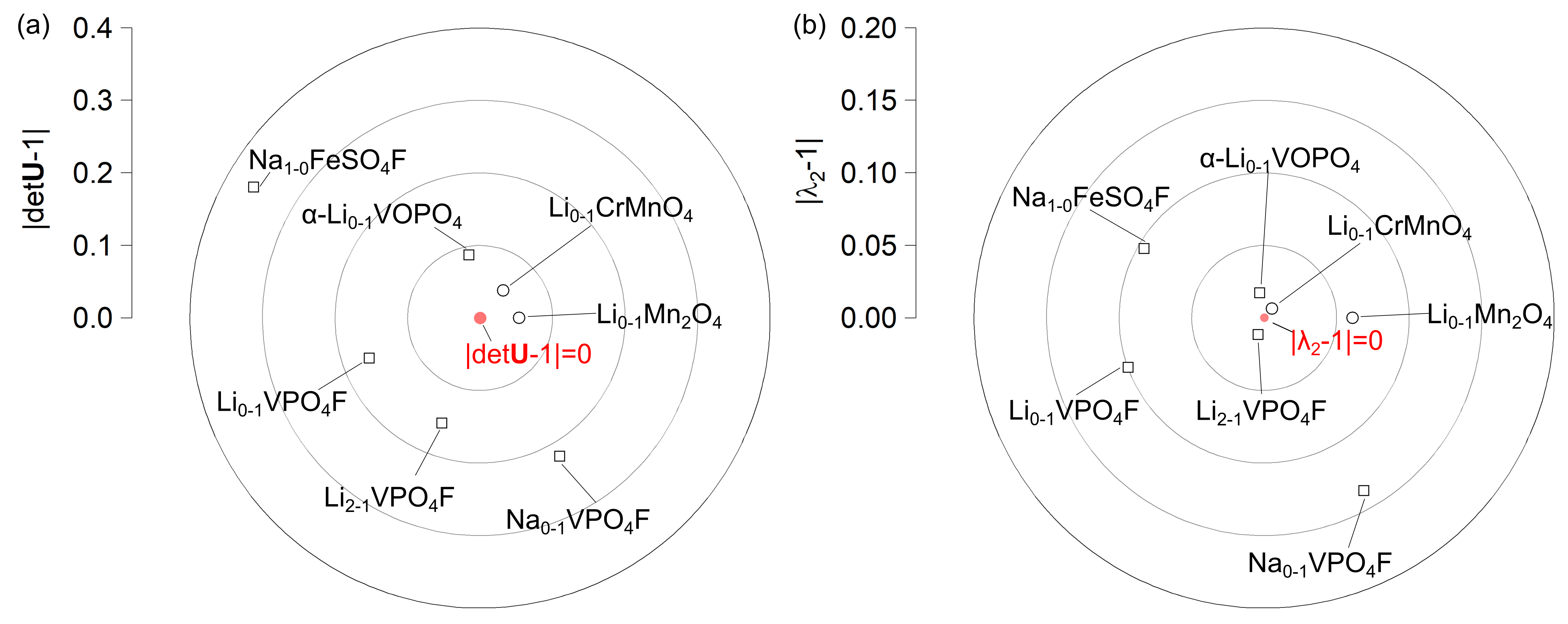}
    \caption{(a) A polar plot illustrating the deviation of the volume changes of representative intercalation compounds from the geometric center (identified by $|\mathrm{det}\mathbf{U} - 1| = 0$ condition). Please note that intercalation materials with a monoclinic reference phase are identified by ``$\square$'' and with other reference phases by ``$\bigcirc$''. \textcolor{black}{For materials with monoclinic reference phase, the geometric center corresponds to $|\mathrm{det}\mathbf{M}-1|=0$ \cite{bhattacharya1992self}.} (b) A polar plot illustrating the deviation of the middle eigenvalues of representative intercalation compounds from the geometric center (identified by $|\lambda_2 - 1| = 0$ condition).}
    \label{fig:special microstructures}
\end{figure}

We next compute the stretch tensors for each intercalation compound in Supplementary Table~7 ($n=25$) and examine whether they form special microstructures, such as the self-accommodation and the $\lambda_2 = 1$ microstructure. 

From Table~\ref{Table1} in Methods, we recall that self-accommodation microstructures must satisfy (a) volume preserving structural transformation, i.e., $|\mathrm{det}\mathbf{U}-1|=0$, and (b) have non-zero stretch along $c-$axis of the lattice during transformation. Fig.~\ref{fig:special microstructures}(a) is a polar plot illustrating the volume change of intercalation compounds, $|\mathrm{det}\mathbf{U}-1|$, from the geometric center at which $|\mathrm{det}\mathbf{U}-1|=0$. The majority of the intercalation compounds do not preserve their volume during intercalation; but a few compounds, such as $\alpha$-$\mathrm{Li_\mathit{x}VOPO_4}$, $\mathrm{Li_\mathit{x}CrMnO_4}$, $\mathrm{Li_\mathit{x}Mn_2O_4}$, have negligible volume changes, $|\mathrm{det}\mathbf{U}-1| \leq 0.05$. Although these compounds have small volume changes, they do not satisfy the remaining geometric conditions to form self-accommodating microstructures, see Supplementary Table~9. However, these compounds serve as suitable candidates that could be systematically doped to satisfy the geometric constraints for forming self-accommodating microstructures during intercalation.

From Table~\ref{Table1}, we note that an exact interface forms between the austenite phase and one variant of the martensite phase when the middle eigenvalue is $\lambda_2 = 1$. We compute the middle eigenvalues for each intercalation compound and estimate the distance $|\lambda_2 - 1|$. Fig.~\ref{fig:special microstructures}(b) is a polar plot illustrating the radial distance of each intercalation compound from the geometric center at which $|\lambda_2 - 1| = 0$. \textcolor{black}{Six of the twenty-five intercalation compounds, including $\alpha-\mathrm{Li_\mathit{x}VOPO_4}$ and $\mathrm{Li_\mathit{x}CrMnO_4}$, lie close to the geometric center with $|\lambda_2 - 1| \leq 0.1$.} \footnote{In addition to the $\lambda_2=1$ condition, we examine whether these intercalation compounds satisfy the remaining cofactor conditions necessary to form highly-reversible microstructures \cite{chen2013study}, see Supplementary Table~10. We note that none of the compounds precisely satisfy these strong compatibility conditions.}

Overall, Figs.~\ref{fig:special microstructures}(a-b) show that none of the intercalation compounds from our representative set exactly satisfy the geometric constraints for self-accommodation and/or $\lambda_2=1$ condition. A few compounds (e.g., \textcolor{black}{$\alpha$-$\mathrm{Li_\mathit{x}VOPO_4}$, $\mathrm{Li_\mathit{x}CrMnO_4}$, and $\mathrm{Li_\mathit{x}Mn_2O_4}$}), however, approximately satisfy the geometric constraints and would serve as candidate materials which can be systematically doped to precisely satisfy the self-accommodation and/or $\lambda_2 = 1$ conditions. These materials would be starting compounds for first-principles calculations or topochemical synthesis that can be atomically engineered leading to novel intercalation cathodes with low hysteresis and high reversibility. 

\subsection*{Study 2}
We next apply our framework to analyze the structural transformations of $n = 5,142$ pairs of intercalation compounds in the Materials Project database.\footnote{We omit 16 intercalation compounds from our microstructural analysis due to significant errors ($\geq 6\%$) in reporting unit cell volume changes on the Materials Project database \cite{Jain2013, jain2011high}.} The Materials Project is an open-access database documenting the structural properties of known and computationally predicted intercalation compounds. We compute the stretch tensor for each pair of these intercalation compounds and analyze whether they undergo lattice-symmetry change during transformation and form self-accommodating and $\lambda_2 = 1$ microstructures. Finally, we compare our microstructural analysis with experimental reports on intercalation material performance (e.g., capacity retention, number of cycles). The findings from our analysis identify select candidate compounds that can be systematically doped to form shape-memory-like microstructures. More generally, our analysis shows a direct link between structural transformation, microstructures, and material performance.

\subsubsection*{Symmetry-lowering transformation}
\begin{figure}[H]
    \centering
    \includegraphics[width=\textwidth]{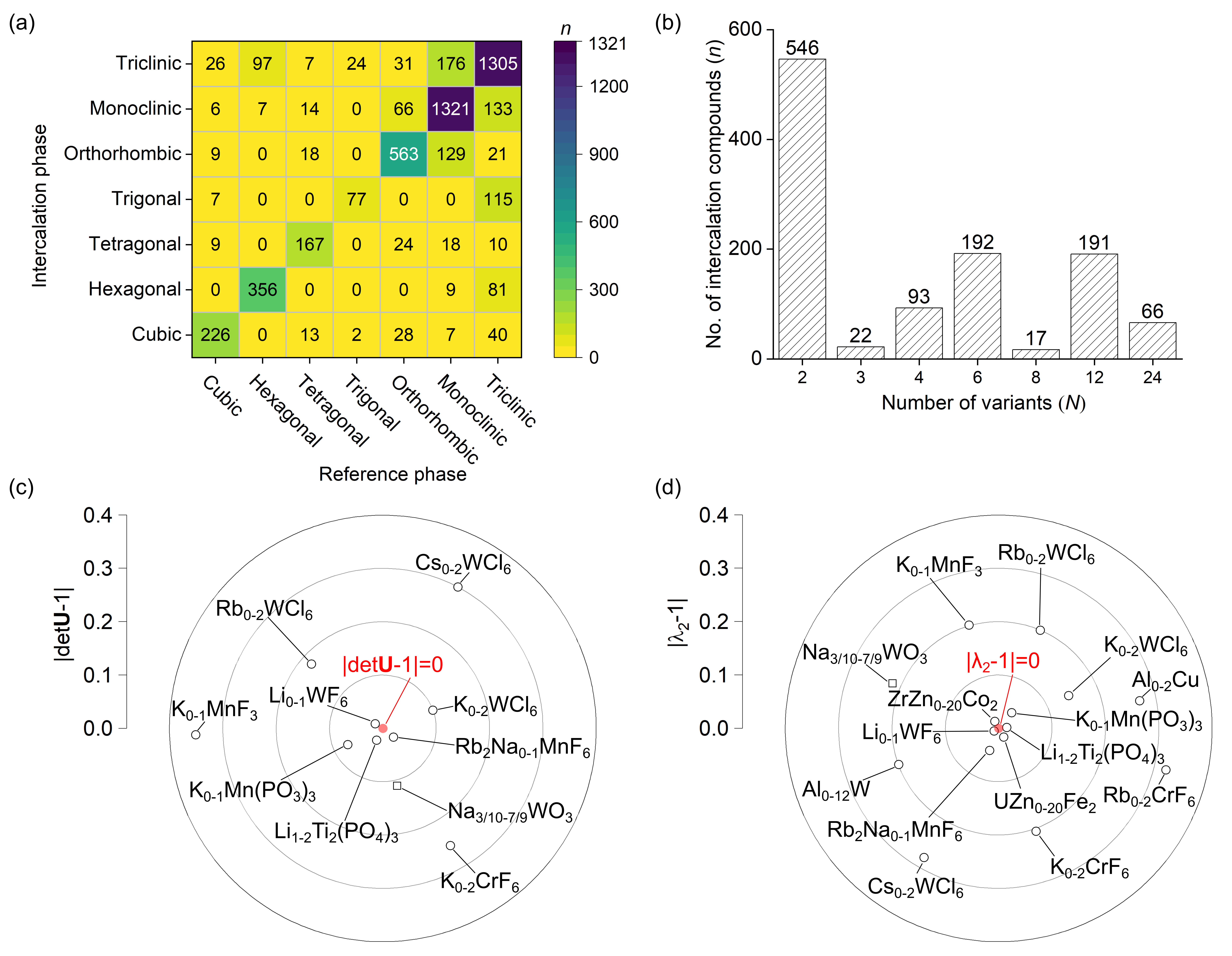}
    \caption{(a) A heat map showing the total number of intercalation compounds $n$ that undergo a lattice-symmetry change during intercalation. Our analysis shows that $78\%$ of the compounds documented in the Materials Project database do not change their lattice symmetry on intercalation. These compounds are shown on the diagonal of the heat map. However, fewer than $22\%$ of the compounds undergo a lattice-symmetry change during transformation (e.g., $n=9$ compounds undergo cubic (reference phase) to tetragonal (intercalation phase) transformation on intercalation) and are shown at corresponding positions that are away from the diagonal. (b) A histogram of the total number of variants $N$ generated during the phase transformation. If a compound undergoes symmetry change during phase transformation, it generates multiple $N\geq2$ variants. Specific details to compute the number of variants are described in the \textcolor{black}{Supplementary Information}. Polar plots illustrating (c) the deviation of the volume changes of intercalation compounds from the geometric center (identified by $|\mathrm{det}\mathbf{U} - 1| = 0$ condition), and (d) the deviation of the middle eigenvalues of intercalation compounds from the geometric center (identified by $|\lambda_2 - 1| = 0$ condition). \textcolor{black}{Please note that intercalation materials with a monoclinic reference phase are identified by ``$\square$'' and with other reference phases by ``$\bigcirc$''.}}
    \label{Fig8}
\end{figure}

Our microstructural analysis shows that only $\sim 22\%$ ($n = 1,127$) of the known intercalation compounds undergo a lattice symmetry change during transformation, see Fig.~\ref{Fig8}(a-b). This lattice symmetry change during transformation is necessary to generate two or more variants that, in turn, are necessary to form shape-memory-like microstructures. 

Fig.~\ref{Fig8}(a) is a heat map showing the number of intercalation compounds that undergo a lattice symmetry change during transformation. We find that majority of the intercalation compounds ($n=4,015$) do not change their lattice symmetries when transformed between the reference and intercalation phases and are concentrated at the diagonal of the heat map in Fig.~\ref{Fig8}(a). These intercalation compounds commonly undergo dilational strains which change the size of a lattice but not its symmetry. Intercalation compounds that undergo a change in lattice symmetry are located away from the diagonal of the heat map in Fig.~\ref{Fig8}(a).

Fig.~\ref{Fig8}(b) is a histogram that shows how many intercalation compounds generate two or more variants during transformation.\footnote{The number of variants $N$ generated during transformation is defined as the ratio of the number of rotations in the point group of Bravais lattices. For example the cubic (24 rotations) $\Leftrightarrow$ tetragonal (8 rotations) transformation in $\mathrm{Li_{0-2}Mn_3CrO_8}$ intercalation compounds generates $N = 24/8 = 3$ variants.} From the $n=1,127$ intercalation compounds undergoing symmetry-lowering transformation, nearly $50\%$ of them generate two variants. For example, the tetragonal $\Leftrightarrow$ orthorhombic transformation in $\mathrm{Li_{0-1}MoO_2}$ generates $N = 2$ variants. Other structural transformations, such as the cubic $\Leftrightarrow$ triclinic in $\mathrm{Li_{0.5-1}NiO_2}$  generates $N =24$ variants. However, these transformations, which generate multiple variants, are rare and correspond to only $\sim 1 \%$ of the intercalation compounds. These compounds that generate multiple variants $N$ during transformation have a greater number of twin solutions and would offer increased flexibility to form shape-memory-like microstructures. Overall, Figs.~\ref{Fig8}(a-b) show that majority of the intercalation materials do not undergo symmetry-lowering transformations and thus do not generate multiple variants. These insights help explain the absence of shape-memory-like microstructures in intercalation materials.

\subsubsection*{Shape-memory-like microstructures}
We note that many intercalation compounds undergoing symmetry lowering transformations in Fig.~\ref{Fig8}(b) decompose into smaller products on cycling (e.g., $\mathrm{Mn_3CrO_8} \to \mathrm{CrO}_2 + \mathrm{MnO}_2$), making them unsuitable for forming shape-memory-like microstructures.\footnote{The intercalation compounds documented on Materials Project have been predicted using high-throughput computing, and the chemical stability of these compounds is often not ascertained using experiments \cite{xiao2019computational,jun2022lithium}. For our analysis, we screen intercalation compounds that are stable (i.e., do not decompose into smaller compounds) and analyze their structural transformation pathways.} However, a smaller set of compounds ($n = 15$ pairs of compounds) do not decompose on intercalation and undergo displacive structural transformations. Our microstructural analysis of these select compounds identifies potential candidates that approximately satisfy the crystallographic design rules for forming self-accommodating and/or $\lambda_2 = 1$ microstructures.

Fig.~\ref{Fig8}(c) is a polar plot that shows the volume change of an intercalation compound, quantified by $|\mathrm{det}\mathbf{U}-1|$, during transformation. Intercalation compounds that form self-accommodation microstructures satisfy the crystallographic design rule $|\mathrm{det}\mathbf{U}-1| = 0$. This condition represents a zero volume change during transformation and corresponds to the geometric center of the polar plot. Fig.~\ref{Fig8}(c) shows that majority of the intercalation compounds, such as the \textcolor{black}{Cs$_{0-2}$WCl$_6$} undergoes a large volume change $\sim30\%$ on intercalation (i.e., $|\mathrm{det}\mathbf{U}-1|=0.3$), and do not satisfy the condition for self-accommodation. However, select compounds, such as the \textcolor{black}{Li$_{1-2}$Ti$_2$(PO$_4$)$_3$} from the NASICON family, \textcolor{black}{$\mathrm{K_{0-1}Mn(PO_3)_3}$} from the Phosphate family, and \textcolor{black}{$\mathrm{Rb_2Na_{0-1}MnF_6}$} from the Fluorite family have negligible volume changes with $|\mathrm{det}\mathbf{U}-1| \leq 0.07$, and approximately satisfy the crystallographic design rule for self-accommodation. These compounds would serve as promising candidates that can be systematically doped to exactly satisfy $|\mathrm{det}\mathbf{U}-1|=0$ \cite{zhang2022DFT}.

Fig.~\ref{Fig8}(d) is a polar plot that shows the middle eigenvalues $\lambda_2$ of intercalation compounds during transformation. The middle eigenvalue $\lambda_2$ quantifies the relative stretch of lattices during transformation, and the crystallographic design rule $\lambda_2=1$ corresponds to microstructures with exactly compatible and stress-free phase-boundaries. Similar to the case in Fig.~\ref{Fig8}(c), our analysis shows that many intercalation compounds do not exactly satisfy the $\lambda_2=1$ condition. However, 6 out of 15 intercalation compounds, including $\mathrm{Li_{1-2}Ti_2(PO_4)_3}$ and $\mathrm{K_{0-1}Mn(PO_3)_3}$, lie close to the geometric center with $|\lambda_2 - 1| \leq 0.04$. \textcolor{black}{This finding is in line with the emergence of $\mathrm{LiTi_2(PO_4)_3}$ as a robust cathode coating material for solid-state batteries \cite{xiao2019computational}}. These results once again show that intercalation compounds in the NASICON and Phosphate families are potential candidates that can be crystallographically engineered to satisfy the geometric constraints to form $\lambda_2 = 1$ microstructures during intercalation. These materials could be doped using techniques, such as the site-selective topochemical synthesis, and is a subject of ongoing research \cite{parker2022alloy}.

\subsubsection*{Comparison with experiments}
We next compare our analytical results with previously published experimental literature on intercalation cathodes \textcolor{black}{(Refs.~\cite{boyadzhieva2015competitive,chen2014pyrophosphate,chen2021role,kim2015anomalous,oh2012reversible,patoux2002lithium,pearce2017evidence,recham20103,sathiya2013reversible,shaju2008stoichiometric,thackeray1997manganese,wang2005improving,wang2015anti,yin2020structural,zhuo2006preparation})}. Specifically, we examine the effect of geometric constraints (e.g., self-accommodation quantified by $|\mathrm{det}\mathbf{U}-1|$ and stress-free phase boundaries quantified by $|\lambda_2 - 1|$) on the performance of intercalation materials (e.g., capacity retention). We find that intercalation cathodes approximately satisfying the crystallographic design rules show improved capacity retention and reversibility.

\begin{figure}[H]
    \centering
    \includegraphics[width=\textwidth]{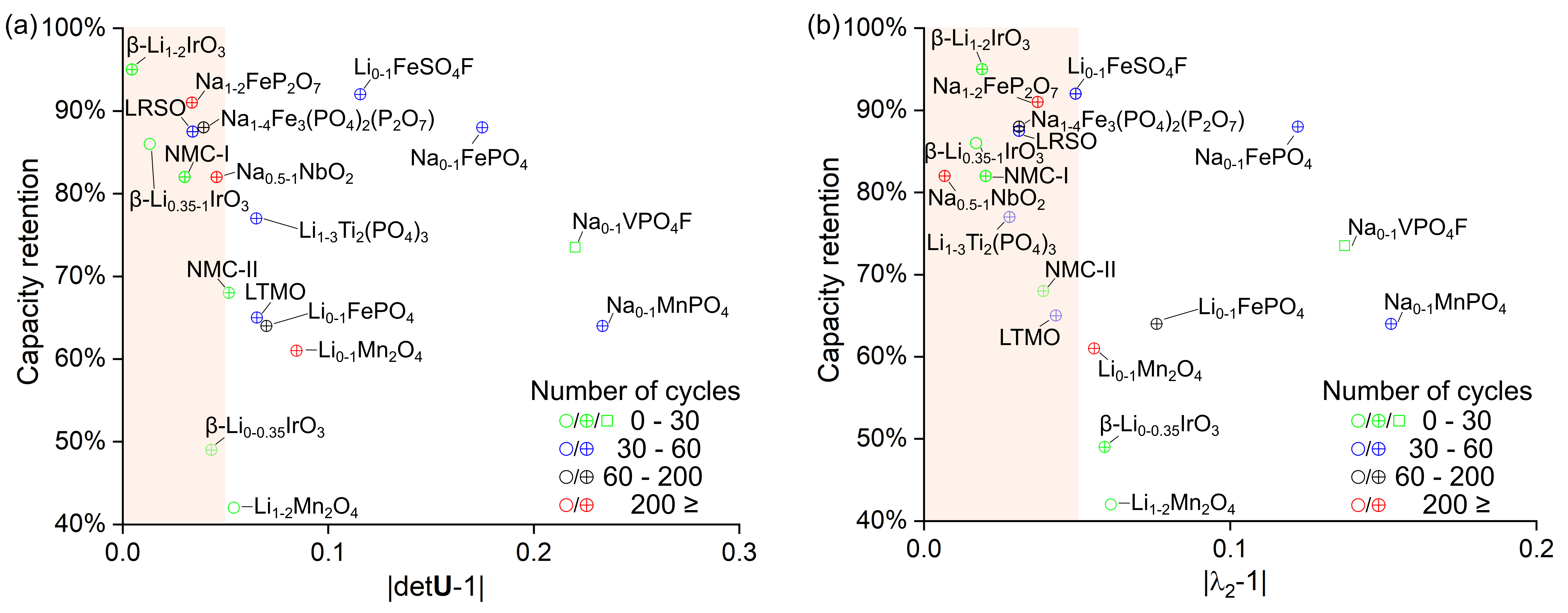}
    \caption{Capacity retention of intercalation cathodes as a function of (a) self-accommodation geometric constraint $|\mathrm{det}\mathbf{U}-1|$ and (b) highly-reversible geometric constraint $|\lambda_2-1|$. Intercalation materials that closely satisfy the geometric constraints, i.e., $|\mathrm{det}\mathbf{U}-1| \to 0$ and $|\lambda_2-1| \to 0$ show higher capacity retention. In these plots, we denote materials with a monoclinic reference phase by ``$\square$'' and other reference phases by ``$\bigcirc$''. Materials that do not show a change in lattice symmetry during intercalation are identified by ``$\oplus$''.}
    \label{Fig9}
\end{figure}

Fig.~\ref{Fig9}(a-b), respectively, show the capacity retention of commonly used intercalation cathodes as a function of the self-accommodation constraint $|\mathrm{det}\mathbf{U} - 1|$ and the $\lambda_2$ microstructure constraint $|\lambda_{2} - 1|$. We note that intercalation cathodes that approximately satisfy the geometric constraints $|\mathrm{det}\mathbf{U} - 1| \to 0$ and $|\lambda_2 - 1| \to 0$ have greater capacity retention. \textcolor{black}{For example, intercalation cathodes such as $\mathrm{\beta-Li_{1-2}IrO_3}$ with $|\mathrm{det}\mathbf{U} - 1| = 0.0045$ and $|\lambda_2-1| = 0.0190$ have 95\% capacity retention even after cycling them for over 30 times \cite{pearce2017evidence}. By contrast, intercalation cathodes such as $\mathrm{Na_{0-1}MnPO_4}$ with $|\mathrm{det}\mathbf{U} - 1| = 0.2334$ and $|\lambda_2-1| = 0.1525$ show only 64\% capacity retention after $\sim20$ charge/discharge cycles \cite{boyadzhieva2015competitive}.} Moreover, Fig.~\ref{Fig9}(a-b) shows that compounds with $\mathrm{det|}\mathbf{U} - 1| \to 0$ and $|\lambda_{2} - 1| \to 0$ not only show increased capacity retention, but are also likely to be cycled for several hundred times (e.g., $\mathrm{Na_{1-2}FeP_2O_7}$ and $\mathrm{Na_{0.5-1}NbO_2}$ are cycled 1000 and 600 times, respectively). These examples show that structural transformations of individual lattices, in addition to the thermodynamic and kinetic driving forces, could affect the performance of intercalation materials.

\textcolor{black}{Fig.~\ref{Fig9}(a-b) shows a few outlier compounds. For example, $\mathrm{Na_{0-1}FePO_4}$ and $\mathrm{Li_{0-1}FeSO_4F}$ have good capacity retention despite having $>10\%$ volume changes \cite{xiang2017accommodating, oh2012reversible, recham20103}. In $\mathrm{Na_{0-1}FePO_4}$ the large volume changes on intercalation amorphizes a fraction of the crystalline material. This amorphous region acts as a buffer for the nanocrystalline $\mathrm{Na_{0-1}FePO_4}$ particles and contributes to increased capacity retention despite the large volume changes in $\mathrm{Na_{0-1}FePO_4}$ nanocrystals \cite{xiang2017accommodating}. Other outlier compounds include $\mathrm{\beta-Li_{0-0.35}IrO_3}$ and NMC-II, which closely satisfy the crystallographic design rules, and yet have lowered capacity retention \cite{pearce2017evidence, yin2020structural}. In these examples, other electrochemical factors such as electrolyte decomposition\cite{pearce2017evidence}, and poor reversibility \cite{yin2020structural} are found to be dominant factors that affect the cycling stability of the material.}

Overall, Fig.~\ref{Fig9}(a-b) shows that intercalation cathodes that approximately satisfy the crystallographic design rules show improved capacity retention and can be reversibly cycled multiple times. These results further emphasize that the structural transformation of individual lattices, in addition to electrode composition, thermodynamic, and kinetic properties, is an important parameter in the design of intercalation materials.

\section*{Discussion}
Our microstructural analysis shows that only a small percentage of known intercalation compounds form twins and austenite/martensite microstructures during intercalation and, a vast majority of the compounds do not satisfy the design principles for self-accommodation and/or $\lambda_2 = 1$ microstructures. This was the case in Study 1, in which selected compounds from the Spinel family form twins and austenite/martensite microstructures; however, their lattice geometries did not exactly satisfy the geometric constraints for self-accommodation and/or $\lambda_2 = 1$ microstructures. In Study 2, less than 22\% of intercalation compounds reported on the Materials Project database undergo symmetry lowering transformation and thus form twins; however, none of these materials precisely satisfy the geometric constraints for self-accommodation and/or $\lambda_2 = 1$ microstructures. These compounds are promising candidates that can be systematically doped to satisfy precise lattice geometries. In the remainder of this section, we discuss some limitations of our findings, compare our results with prior work on understanding intercalation material reversibility, and present the potential impact of our work on the materials discovery program.

Two features of this work limit the conclusions we can draw from our microstructural analysis. First, we minimize a distance function $||\mathbf{U}^{-2}-\mathbf{I}||^2$ to identify an optimal stretch tensor $\mathbf{U}$. This distance function quantifies the total strain of a structural transformation; however, minimizing this function can generate more than one solution for the stretch tensor. In these cases, we choose the stretch tensor describing a lattice-symmetry lowering transformation as this would generate energy minimizing microstructures. Second, the accuracy of the computed structural data of intercalation compounds on the Materials Project database affects our microstructural analysis. For example, we identify candidate compounds that approximately satisfy the design constraints for self-accommodation and/or $\lambda_2 = 1$ microstructures; however, we have limited information on the chemical stability of these compounds. Many of these compounds were predicted using first-principle calculations, and rigorous experimental investigations would be necessary to ascertain their chemical stability. With these reservations in mind, we next discuss the impact of our findings on the materials discovery program.

A key feature of our work is that we establish a theoretical framework and crystallographic design rules to guide the discovery of intercalation materials with minimum volume changes and interfacial stresses. For example, our results identify intercalation compounds that approximately satisfy the design rules necessary to form self-accommodation and/or $\lambda_2 = 1$ microstructures. These compounds serve as potential candidates that can be systematically engineered (e.g., using site-selective topochemical synthesis) to satisfy precise lattice geometries. This theory-guided search for intercalation compounds would accelerate the discovery of novel intercalation materials with improved reversibility.

Another significant feature of our work is that we show a direct link between structural transformations, microstructures, and material behavior. For example, we compute the stretch tensor for Li$_{1-2}$Mn$_2$O$_4$ and quantitatively predict the microstructural features measured in Fig.~\ref{Fig6}. Furthermore, our analysis in Fig.~\ref{Fig9} shows that intercalation compounds that approximately satisfy the self-accommodating (i.e., $|\mathrm{det}\mathbf{U} - 1| = 0$) and $\lambda_2 = 1$ design constraints have enhanced ($\geq 80\%$) capacity retention. \textcolor{black}{In another example, we show compounds such as $\mathrm{Na_{1-2}FeP_2O_7}$ that closely satisfy the design constraints $|\mathrm{det}\mathbf{U}-1| \to 0$ and $|\lambda_2 - 1| \to 0$ are also highly reversible ($>1000$ times).} These examples demonstrate that structural transformation pathways of individual lattices, in addition to the electrochemical operating conditions and diffusion kinetics, play an important role in material performance and reversibility.

\textcolor{black}{To summarize}, we quantify structural transformations in intercalation compounds and establish crystallographic design rules necessary to form shape-memory-like microstructures in intercalation materials. Our findings show that majority of the known intercalation compounds do not satisfy the crystallographic design rules necessary to form microstructures that are self-accommodating and/or have stress-free interfaces (i.e., $\lambda_2 = 1$ microstructure). However, we identify candidate intercalation compounds--such as the spinel Li$_x$Mn$_2$O$_4$ and NASICON Li$_x$Ti$_2$(PO$_4$)$_3$--that approximately satisfy the crystallographic design rules. These compounds serve as promising candidates that can be systematically doped to satisfy precise lattice geometries and thus form microstructures with minimum volume changes and/or stress-free interfaces. More generally, our analysis and the crystallographic design principles serve as a theoretical guide to discovering a new generation of intercalation materials with reduced volume changes and stress-free microstructures.

\section*{Methods}
\subsection*{Crystallographic Design Principles}
\begin{figure}[H]
    \centering
    \includegraphics[width=\textwidth]{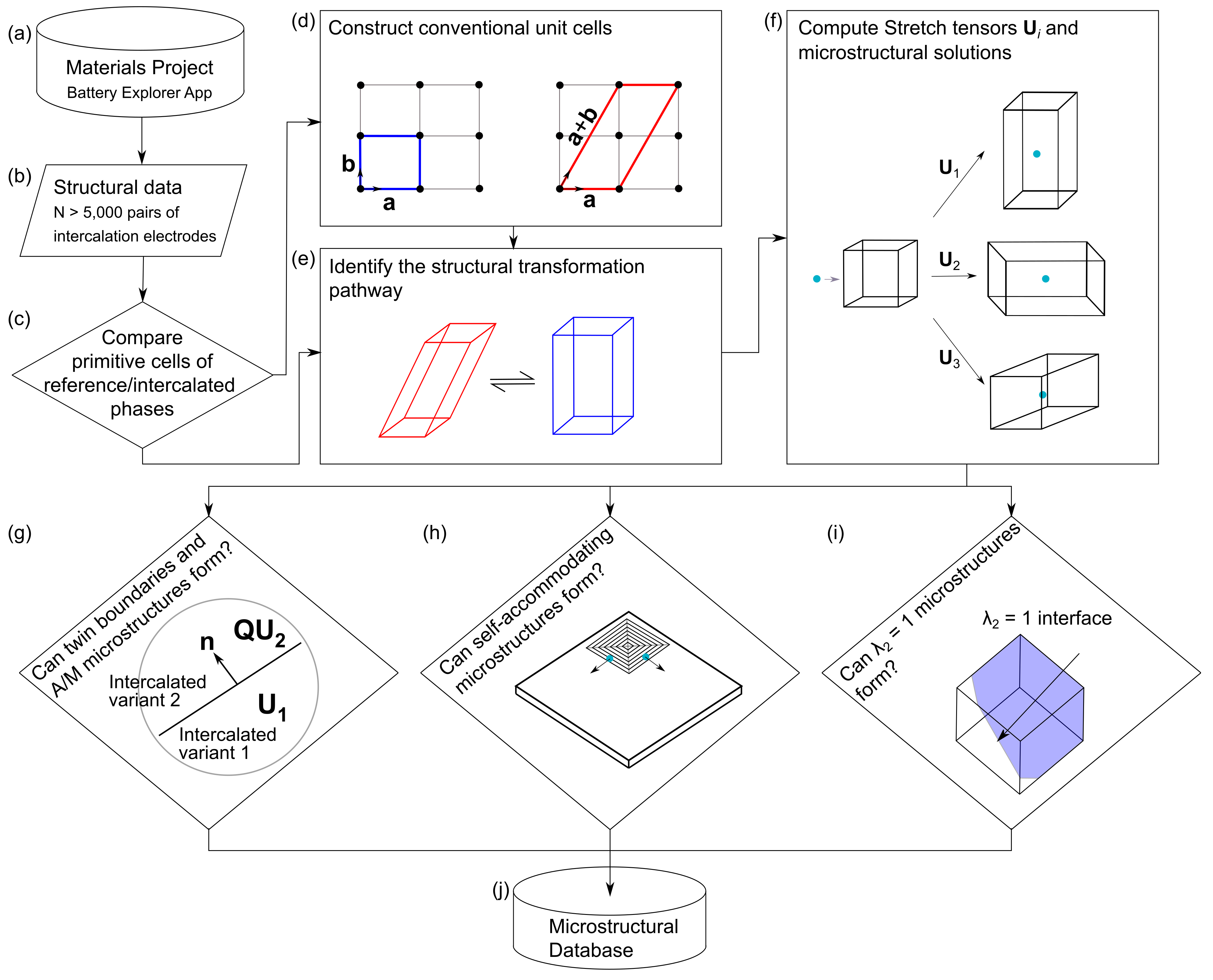}
    \caption{Workflow steps of our theoretical framework. (a-b) We first extract structural data of over 5,000 pairs of intercalation compounds from the Materials Project database. We then apply our developed structural transformation algorithm (c-d) to determine unit cells in the reference and intercalated phases, and (e-f) construct the optimal stretch tensor to describe the transformation pathway. Finally, we use the computed stretch tensor for each compound to identify potential candidates that can form (g) twin boundaries, austenite/martensite (A/M), (h) self-accommodating, or (i) highly reversible microstructures. (j) The solutions of these compounds are stored in the microstructural database.}
    \label{Fig3}
\end{figure}

In this section, we outline the theoretical framework that we use to compute the structural transformation of unit cells during intercalation and to analyze how these transformations collectively generate microstructures at the continuum scale, see Fig.~\ref{Fig3}. Before we outline our framework, we describe a Cauchy-Born rule that is central to our analysis: The Cauchy-Born rule is a basic hypothesis used in the mathematical formulation of solid mechanics which relates the deformation of the bulk solid to the movement of atoms in a crystal \cite{ericksen2008cauchy}. This rule gives an exact correspondence between the continuum microstructures at a material point and the structural transformations of individual lattices. This Cauchy-Born rule allows us to impose special lattice geometries as a necessary condition for characteristic microstructures (e.g., austenite-martensite, self-accommodating, $\lambda_2=1$) to form during phase transformations.\footnote{Other factors, such as material thermodynamics and diffusion kinetics could affect microstructural evolution; however, none of these factors are known to be as important as the specific lattice geometries of the material \cite{james2005way}.} We use the Cauchy-Born rule in our theoretical framework and establish the crystallographic design principles necessary to form shape-memory-like microstructures in intercalation compounds.

\subsubsection*{Structural Transformation}

\begin{figure}[H]
    \centering
    \includegraphics[width=\textwidth]{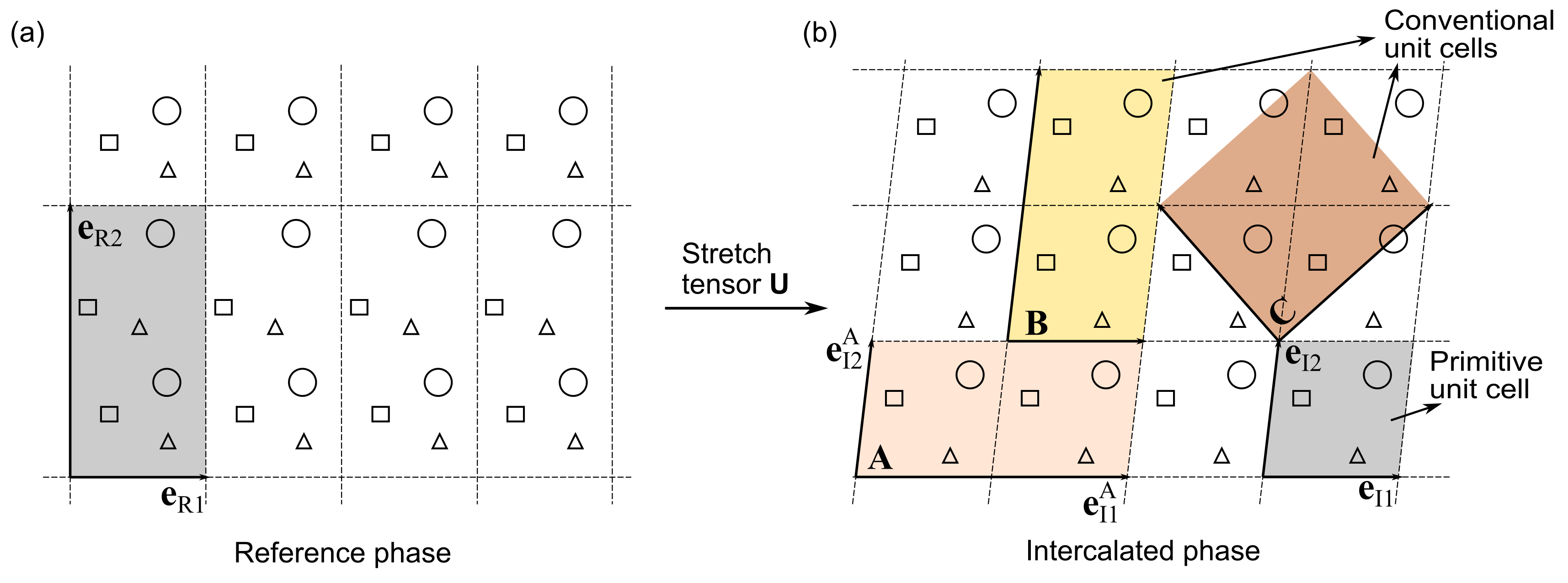}
    \caption{A schematic illustration of primitive and conventional unit cells in reference and intercalated (or transformed) phases. (a) Vectors $\{\mathbf{e}_\mathrm{R1}, \mathbf{e}_\mathrm{R2}\}$ enclose a smallest repeatable unit cell (or the primitive cell) in the reference phase. (b) Vectors $\{\mathbf{e}_\mathrm{I1}, \mathbf{e}_\mathrm{I2}\}$ enclose a primitive cell in the intercalated phase. The primitive cell of the reference phase contains twice as many atoms as that of the primitive cell of the intercalated phase. In subfigure (b) We illustrate three representative examples of conventional unit cells--$\mathrm{A, B, C}$--which contain the same number of atoms as the primitive cell of the reference phase. We note that these conventional cells can be described in multiple ways, and thus the stretch tensor $\mathbf{U}$ mapping unit cells between reference and intercalated phases can have more than one solution.}
    \label{Fig4}
\end{figure}

We describe the structural transformation of unit cells in intercalation compounds using a stretch tensor $\mathbf{U}$.\footnote{A stretch tensor is a mathematical quantity, i.e., a rank-2 matrix, which provides a linear mapping of unit cells between the reference and intercalation phases. A brief description of mathematical quantities introduced in this paper is provided in the \textcolor{black}{Supplementary Table 2}.} This stretch tensor maps a unit cell in the reference phase (i.e., before intercalation) to the corresponding unit cell in the intercalated phase (i.e., after intercalation). This stretch tensor, for a given compound, is not unique and can have multiple solutions based on how we choose the unit cells for reference and intercalated phases, see Fig.~\ref{Fig4}. In this section, we describe our algorithm that determines an optimal stretch tensor for a given intercalation compound. We use these optimal stretch tensors to predict microstructural patterns at the continuum scale.

Fig.~\ref{Fig4} shows a schematic illustration of a unit cell in the reference phase, which is described using three linearly independent lattice vectors $\mathbf{E}_{\mathrm{R}} = \{\mathbf{e}_{\mathrm{R}1}, \mathbf{e}_{\mathrm{R}2}, \mathbf{e}_{\mathrm{R}3}\}$. These lattice vectors enclose the smallest repeatable volume and constitute a primitive unit cell. On transformation the materials undergo a structural change and a unit cell is now described by lattice vectors $\mathbf{E}_{\mathrm{I}} = \{\mathbf{e}_{\mathrm{I}1}, \mathbf{e}_{\mathrm{I}2}, \mathbf{e}_{\mathrm{I}3}\}$. However, this definition of a unit cell is not unique as other combinations of lattice vectors describing unit cells $\mathbf{E}^{\mathrm{A}}_{\mathrm{I}} = \{\mathbf{e}^{\mathrm{A}}_{\mathrm{I1}}, \mathbf{e}^{\mathrm{A}}_{\mathrm{I2}}, \mathbf{e}^{\mathrm{A}}_{\mathrm{I3}}\}$ (or $\mathbf{E}^{\mathrm{B}}_{\mathrm{I}}$, or $\mathbf{E}^{\mathrm{C}}_{\mathrm{I}}$) generate the same lattice points in 3D space, see Fig.~\ref{Fig4}. These latter unit cells, called conventional cells, typically enclose a larger volume (relative to $\mathbf{E}_{\mathrm{I}}$) and can be described in multiple ways. Each conventional cell is related to the primitive unit cell via a lattice correspondence matrix $\mathbf{P} \in \mathbb{Z}^3$ as $\mathbf{E}^{\mathrm{A}}_{\mathrm{I}}=\mathbf{E_{\mathrm{I}}P^{\mathrm{A}}}$, and several correspondence matrices can exist based on our choice of the conventional cells.\footnote{Similarly, the unit cells in the reference phase can be described in multiple ways.} Based on our choice of unit cells for the reference and intercalation phases, we would have more than one stretch tensor describing the transformation between the two phases. 

Following Refs.~\cite{chen2016determination,thomas2021comparing}, we propose an algorithm to identify an optimal stretch tensor, from a list of potential stretch tensors, by minimizing the transformation pathways between the reference and intercalated unit cells. We describe this algorithm in detail in the \textcolor{black}{Supplementary Information}, and outline the three key steps below:

\begin{enumerate}
    \item Determine the primitive unit cells of reference and intercalated phases and identify all potential lattice correspondence matrices $\mathbf{P}$ for a given compound. In the \textcolor{black}{Supplementary Information}, we outline the specific steps to compute $\mathbf{P}$ using a pair of primitive unit cells as an example.
    \item Compute the stretch tensor $\mathbf{U}$ mapping a unit cell of the reference phase to the corresponding unit cell of the intercalated phase. The general form for the mapping is given by $\mathbf{F}\mathbf{E}_\mathrm{R}\mathbf{P}_\mathrm{R} = \mathbf{E}_\mathrm{I}\mathbf{P}_\mathrm{I}$ in which $\mathbf{F}$ is the deformation gradient, and the transformation stretch tensor $\mathbf{U}$ is the unique positive-definite square root of $\mathbf{F}^{\mathrm{T}}\mathbf{F}$.\footnote{\textcolor{black}{Please see Supplementary Table 2 for the physical meaning of a deformation gradient.}} This stretch tensor $\mathbf{U}$ is unique for a given pair of correspondence matrices $\mathbf{P}_\mathrm{R}, \mathbf{P}_\mathrm{I}$, and we determine all possible stretch tensors by cycling through the set of correspondence matrices identified in the first step.
    \item Identify an optimal lattice correspondence (and thus the stretch tensor) that minimizes a distance function representing the total strain of a structural transformation: $\mathrm{dist}(\mathbf{P}_\mathrm{R}, \mathbf{P}_\mathrm{I}, \mathbf{E}_\mathrm{R}, \mathbf{E}_\mathrm{I})=\left\|\mathbf{U}^{-2}-\mathbf{I}\right\|^2$. Here, $\left\|{.}\right\|$ is the Frobenius norm, $\left\|{\mathbf{A}}\right\|^2= \mathbf{A}:\mathbf{A} = \mathrm{Tr}(\mathbf{A}^\mathrm{T}\mathbf{A})$. By minimizing this distance function over the set of stretch tensors computed in the second step, we determine the optimal lattice correspondence and the transformation stretch tensor for a given intercalation compound.
\end{enumerate}

Further details of this algorithm using $\mathrm{Li}_\mathit{x}\mathrm{Fe_2(MoO_4)_3}$ as an example is described in the \textcolor{black}{Supplementary Information}. In principle, this algorithm can be applied to determine stretch tensors for any crystalline material undergoing a first-order phase transformation; however, for the present work, we apply our theoretical framework to intercalation compounds commonly used in batteries. We use the computed optimal stretch tensors in our microstructural analysis---that is, we determine whether an intercalation compound satisfies the crystallographic design principles described in the next section.

\subsubsection*{Microstructures}

The crystallographic theory of martensites is an energy minimization theory \cite{ball1989fine} in which the material's energy is described as a function of a lattice deformation gradient $\mathbf{F}$. The lattice deformation gradient at the $atomic$ scale penalizes the elastic energy arising from lattice misfit. Minimizing this elastic energy then results in the formation of finely twinned microstructures at the \emph{continuum} scale, see Fig.~\ref{Fig2}. This theory has been widely used to explain the characteristic microstructures in shape memory alloys, ferroelectrics, ferromagnets, and more recently in light-interactive materials \cite{james1998magnetostriction, li2005domain, naumov2015mechanically}. We use this theory to establish the geometric conditions necessary to form shape-memory-like microstructures in intercalation materials.

Following the crystallographic theory of martensites, we next list the design principles necessary to form the four candidate shape-memory-like microstructures in Table~\ref{Table1}. These design principles identify specific lattice geometries quantified by stretch tensors $\mathbf{U}$ that are necessary to form twins, austenite/martensite, self-accommodating, and $\lambda_2 = 1$ (stress-free) microstructures. We identify these microstructures to have several advantages (as described below) and that, if stabilized in intercalation materials, would help mitigate their chemo-mechanical degradation:
\begin{enumerate}
    \item Shape-memory-like microstructures, such as the twins and austenite/martensite microstructure, reduce the elastic energy arising from misfit strains between neighboring lattices and across phase boundaries, respectively. This reduced energy suppresses microcracking of intercalation materials with repeated usage. 
    
    \item The self-accommodation microstructure, adapts to the original shape of the material without any macroscopic change in volume, despite significant structural changes at the atomic scale. These microstructures have the potential to eliminate large volume changes of intercalation electrodes, which in turn lead to delamination in the composite electrode/electrolyte system.
    
    \item The $\lambda_2 = 1$ microstructure corresponds to an exactly compatible and stress-free phase boundary, which forms between the reference and intercalated phases. These microstructures eliminate interfacial stresses that commonly arise during phase transformations, and have contributed to a phenomenal improvement in the reversible cycling of phase transformation materials \cite{chluba2015ultralow}. We provide a detailed perspective on these microstructures in the \textcolor{black}{Supplementary Information}. 
\end{enumerate}

\begin{table}[H]
    \small
    \centering
    \renewcommand\arraystretch{1.5}
    \begin{tabular}{llll}
    \hline
        Microstructure & Geometric condition & Notation & Advantages\\
    \hline        
        Twin & $\mathbf{QU_\mathit{I}-U_\mathit{J}=a\otimes\hat{n}}$ & Stretch tensors $\mathbf{U}_I$ $\mathbf{U}_J$ & Exactly compatible\\
        
        & & Rotation matrix $\mathbf{Q}$& interface \\
        
        & & Vectors $\mathbf{a \neq \mathrm{0}}$, $\hat{\mathbf{n}}$ \\
        
        & & & \\
        
        Austenite/Martensite & $\mathbf{Q}'(f\mathbf{QU}_{J}+(1-f)\mathbf{U}_{I})$ & Rotation matrices $\mathbf{Q}',\mathbf{Q}$ & Energy minimizing\\
        
        & $= \mathbf{I}+\mathbf{b}\otimes\mathbf{\hat{m}}$ & Volume fraction $f$ & deformation\\
        
        & & Identity matrix $\mathbf{I}$ & \\
        & & Vectors $\mathbf{b \neq \mathrm{0}}, \hat{\mathbf{m}}$ & \\
        
        & & & \\
        
        Self-accommodating\tablefootnote{The listed self-accommodation condition applies to all Bravais lattice symmetries except the monoclinic symmetry in the reference phase. For these cases, we use det$\mathbf{M}=1$ and further details are described in Supplementary Information.} & det$\mathbf{U}$ = 1 & Stretch tensor $\mathbf{U}$ & No macroscopic change\\
        
        & Non-zero stretch along &  & in shape and volume\\
        & $c$-axis&&\\
        
        & & & \\
        
        $\lambda_2 = 1$ & $\mathbf{QU}_{I} - \mathbf{I} = \mathbf{b} \otimes \mathbf{\hat{m}}$ & Rotation matrix $\mathbf{Q}$ & Stress-free phase\\
        
        & & Vectors $\mathbf{b \neq \mathrm{0}}, \hat{\mathbf{m}}$ & boundary\\
        
    \hline
    
\end{tabular}
\caption{Crystallographic design principles to form twin interfaces, austenite/martensite microstructures, self-accommodating, and $\lambda_2 = 1$ microstructures in intercalation materials. The stretch tensors of a material must satisfy specific geometric conditions to form shape-memory-like microstructures. Further details on the notations used here are listed in \textcolor{black}{Supplementary Table 1}.}
\label{Table1}
\end{table}

\vspace{2mm}
\noindent {\bf Data Availability}. The authors declare that the data supporting the findings of this study are available within the paper and its supplementary information files. 

\vspace{2mm}
\noindent {\bf Acknowledgment}. The authors acknowledge the Center for Advanced Research Computing at the University of Southern California for providing resources that contributed to the research results reported in this paper. A.R.B acknowledges the support of a Provost Assistant Professor Fellowship, Gabilan WiSE fellowship, and USC's start-up funds.

\vspace{2mm}
\noindent {\bf Author contribution}. ARB conceptualized the project, designed the methodology, and procured funding. DZ and ARB worked on model development, theoretical analysis, theoretical calculations, and visualization of data. Both authors were involved in the writing of the paper. 

\vspace{2mm}
\noindent {\bf Competing Interests}. The authors declare that there are no competing interests.

\printbibliography

@article{bhattacharya1992self,
  title={Self-accommodation in martensite},
  author={Bhattacharya, Kaushik},
  journal={Archive for Rational Mechanics and Analysis},
  volume={120},
  number={3},
  pages={201--244},
  year={1992},
  publisher={Springer}
}

@article{erichsen2020tracking,
  title={Tracking the diffusion-controlled lithiation reaction of $\mathrm{LiMn_2O_4}$ by In Situ TEM},
  author={Erichsen, Torben and Pfeiffer, Bjorn and Roddatis, Vladimir and Volkert, Cynthia A},
  journal={ACS Applied Energy Materials},
  volume={3},
  number={6},
  pages={5405--5414},
  year={2020},
  publisher={ACS Publications}
}

@incollection{ball1989fine,
  title={Fine phase mixtures as minimizers of energy},
  author={Ball, John M and James, Richard D},
  booktitle={Analysis and Continuum Mechanics},
  pages={647--686},
  year={1989},
  publisher={Springer}
}

@article{zhang2021film,
  title={Film strains enhance the reversible cycling of intercalation electrodes},
  author={Zhang, Delin and Sheth, Jay and Sheldon, Brian W and Balakrishna, Ananya Renuka},
  journal={Journal of the Mechanics and Physics of Solids},
  volume={155},
  pages={104551},
  year={2021},
  publisher={Elsevier}
}

@article{balakrishna2022crystallographic,
  title={Crystallographic design of intercalation materials},
  author={Balakrishna, Ananya Renuka},
  journal={Journal of Electrochemical Energy Conversion and Storage},
  volume={},
  pages={arXiv:2204.04525},
  year={2022},
  publisher={in press}
}

@article{rudraraju2016mechanochemical,
  title={Mechanochemical spinodal decomposition: a phenomenological theory of phase transformations in multi-component, crystalline solids},
  author={Rudraraju, Shiva and Van der Ven, Anton and Garikipati, Krishna},
  journal={npj Computational Materials},
  volume={2},
  number={1},
  pages={1--9},
  year={2016},
  publisher={Nature Publishing Group}
}

@misc{brock2016international,
  title={International tables for crystallography volume A: Space-group symmetry},
  author={Brock, Carolyn Pratt and Hahn, T and Wondratschek, H and M{\"u}ller, U and Shmueli, U and Prince, E and Authier, A and Kopsk{\`y}, V and Litvin, DB and Arnold, E and others},
  year={2016},
  publisher={Wiley Online Library}
}

@article{chen2016determination,
  title={Determination of the stretch tensor for structural transformations},
  author={Chen, Xian and Song, Yintao and Tamura, Nobumichi and James, Richard D},
  journal={Journal of the Mechanics and Physics of Solids},
  volume={93},
  pages={34--43},
  year={2016},
  publisher={Elsevier}
}

@article{recham20103,
  title={A 3.6 $\mathrm{V}$ lithium-based fluorosulphate insertion positive electrode for lithium-ion batteries},
  author={Recham, Nadir and Chotard, Jean-Noel and Dupont, Loic and Delacourt, Charles and Walker, Wesley and Armand, Michel and Tarascon, Jean-Marie},
  journal={Nature materials},
  volume={9},
  number={1},
  pages={68--74},
  year={2010},
  publisher={Nature Publishing Group}
}

@article{oh2012reversible,
  title={Reversible $\mathrm{NaFePO_4}$ electrode for sodium secondary batteries},
  author={Oh, Seung-Min and Myung, Seung-Taek and Hassoun, Jusef and Scrosati, Bruno and Sun, Yang-Kook},
  journal={Electrochemistry communications},
  volume={22},
  pages={149--152},
  year={2012},
  publisher={Elsevier}
}

@article{wang2015anti,
  title={Anti-$\mathrm{P}$2 structured $\mathrm{Na_{0.5}NbO_2}$ and its negative strain effect},
  author={Wang, Xuefeng and Gao, Yurui and Shen, Xi and Li, Yejing and Kong, Qingyu and Lee, Sungsik and Wang, Zhaoxiang and Yu, Richeng and Hu, Yong-Sheng and Chen, Liquan},
  journal={Energy \& Environmental Science},
  volume={8},
  number={9},
  pages={2753--2759},
  year={2015},
  publisher={Royal Society of Chemistry}
}

@article{wang2005improving,
  title={Improving the rate performance of $\mathrm{LiFePO_4}$ by $\mathrm{Fe}$-site doping},
  author={Wang, Deyu and Li, Hong and Shi, Siqi and Huang, Xuejie and Chen, Liquan},
  journal={Electrochimica Acta},
  volume={50},
  number={14},
  pages={2955--2958},
  year={2005},
  publisher={Elsevier}
}

@article{shaju2008stoichiometric,
  title={A stoichiometric nano-$\mathrm{LiMn_2O_4}$ spinel electrode exhibiting high power and stable cycling},
  author={Shaju, Kuthanapillil M and Bruce, Peter G},
  journal={Chemistry of Materials},
  volume={20},
  number={17},
  pages={5557--5562},
  year={2008},
  publisher={ACS Publications}
}

@article{pearce2017evidence,
  title={Evidence for anionic redox activity in a tridimensional-ordered Li-rich positive electrode $\beta$-$\mathrm{Li_2IrO_3}$},
  author={Pearce, Paul E and Perez, Arnaud J and Rousse, Gwenaelle and Sauban{\`e}re, Mathieu and Batuk, Dmitry and Foix, Dominique and Mccalla, Eric and Abakumov, Artem M and Van Tendeloo, Gustaaf and Doublet, Marie-Liesse and others},
  journal={Nature materials},
  volume={16},
  number={5},
  pages={580--586},
  year={2017},
  publisher={Nature Publishing Group}
}

@article{boyadzhieva2015competitive,
  title={Competitive lithium and sodium intercalation into sodium manganese phospho-olivine $\mathrm{NaMnPO_4}$ covered with carbon black},
  author={Boyadzhieva, T and Koleva, V and Zhecheva, E and Nihtianova, D and Mihaylov, L and Stoyanova, R},
  journal={RSC advances},
  volume={5},
  number={106},
  pages={87694--87705},
  year={2015},
  publisher={Royal Society of Chemistry}
}

@article{jain2011high,
  title={A high-throughput infrastructure for density functional theory calculations},
  author={Jain, Anubhav and Hautier, Geoffroy and Moore, Charles J and Ong, Shyue Ping and Fischer, Christopher C and Mueller, Tim and Persson, Kristin A and Ceder, Gerbrand},
  journal={Computational Materials Science},
  volume={50},
  number={8},
  pages={2295--2310},
  year={2011},
  publisher={Elsevier}
}

@article{Jain2013,
author = {Jain, Anubhav and Ong, Shyue Ping and Hautier, Geoffroy and Chen, Wei and Richards, William Davidson and Dacek, Stephen and Cholia, Shreyas and Gunter, Dan and Skinner, David and Ceder, Gerbrand and Persson, Kristin a.},
doi = {10.1063/1.4812323},
issn = {2166532X},
journal = {APL Materials},
number = {1},
pages = {011002},
title = {{Commentary: The Materials Project: A materials genome approach to accelerating materials innovation}},
url = {https://doi.org/10.1063/1.4812323},
volume = {1},
year = {2013}
}

@article{sathiya2013reversible,
  title={Reversible anionic redox chemistry in high-capacity layered-oxide electrodes},
  author={Sathiya, Mariyappan and Rousse, Gwena{\"e}lle and Ramesha, K and Laisa, CP and Vezin, Herv{\'e} and Sougrati, Moulay Tahar and Doublet, Marie-Liesse and Foix, D and Gonbeau, Danielle and Walker, W and others},
  journal={Nature materials},
  volume={12},
  number={9},
  pages={827--835},
  year={2013},
  publisher={Nature Publishing Group}
}

@article{chen2021role,
  title={Role of fluorine in chemomechanics of cation-disordered rocksalt cathodes},
  author={Chen, Dongchang and Zhang, Jin and Jiang, Zhisen and Wei, Chenxi and Burns, Jordan and Li, Linze and Wang, Chongmin and Persson, Kristin and Liu, Yijin and Chen, Guoying},
  journal={Chemistry of Materials},
  volume={33},
  number={17},
  pages={7028--7038},
  year={2021},
  publisher={ACS Publications}
}

@article{dresselhaus1981intercalation,
  title={Intercalation compounds of graphite},
  author={Dresselhaus, MS and Dresselhaus, G},
  journal={Advances in Physics},
  volume={30},
  number={2},
  pages={139--326},
  year={1981},
  publisher={Taylor \& Francis}
}

@article{whittingham1978chemistry,
  title={Chemistry of intercalation compounds: Metal guests in chalcogenide hosts},
  author={Whittingham, M Stanley},
  journal={Progress in Solid State Chemistry},
  volume={12},
  number={1},
  pages={41--99},
  year={1978},
  publisher={Elsevier}
}

@article{padhi1997phospho,
  title={Phospho-olivines as positive-electrode materials for rechargeable lithium batteries},
  author={Padhi, Akshaya K and Nanjundaswamy, Kirakodu S and Goodenough, John B},
  journal={Journal of the electrochemical society},
  volume={144},
  number={4},
  pages={1188},
  year={1997},
  publisher={IOP Publishing}
}

@article{liu2020spontaneous,
  title={Spontaneous self-intercalation of copper atoms into transition metal dichalcogenides},
  author={Liu, Xiao-Chen and Zhao, Shuyang and Sun, Xueping and Deng, Liangzi and Zou, Xiaolong and Hu, Youcheng and Wang, Yun-Xiao and Chu, Ching-Wu and Li, Jia and Wu, Jingjie and others},
  journal={Science advances},
  volume={6},
  number={7},
  pages={eaay4092},
  year={2020},
  publisher={American Association for the Advancement of Science}
}

@article{zhou2014two,
  title={Two-phase transition of $\mathrm{Li}$-intercalation compounds in $\mathrm{Li}$-ion batteries},
  author={Zhou, Haoshen and others},
  journal={Materials Today},
  volume={17},
  number={9},
  pages={451--463},
  year={2014},
  publisher={Elsevier}
}

@article{lim2016intercalation,
  title={Intercalation of solid hydrogen into graphite under pressures},
  author={Lim, Jinhyuk and Yoo, Choong-Shik},
  journal={Applied Physics Letters},
  volume={109},
  number={5},
  pages={051905},
  year={2016},
  publisher={AIP Publishing LLC}
}

@article{nadkarni2019modeling,
  title={Modeling the metal--insulator phase transition in $\mathrm{Li_\mathit{x}CoO_2}$ for energy and information storage},
  author={Nadkarni, Neel and Zhou, Tingtao and Fraggedakis, Dimitrios and Gao, Tao and Bazant, Martin Z},
  journal={Advanced Functional Materials},
  volume={29},
  number={40},
  pages={1902821},
  year={2019},
  publisher={Wiley Online Library}
}

@article{lewis2019chemo,
  title={Chemo-mechanical challenges in solid-state batteries},
  author={Lewis, John A and Tippens, Jared and Cortes, Francisco Javier Quintero and McDowell, Matthew T},
  journal={Trends in Chemistry},
  volume={1},
  number={9},
  pages={845--857},
  year={2019},
  publisher={Elsevier}
}

@article{zhang2020stress,
  title={Stress-Induced Intercalation Instability},
  author={Zhang, Youtian and Tang, Ming},
  journal={Acta Materialia},
  volume={201},
  pages={158--166},
  year={2020},
  publisher={Elsevier}
}

@article{chen2006electron,
  title={Electron microscopy study of the $\mathrm{LiFePO_4}$ to $\mathrm{FePO_4}$ phase transition},
  author={Chen, Guoying and Song, Xiangyun and Richardson, Thomas J},
  journal={Electrochemical and Solid-state letters},
  volume={9},
  number={6},
  pages={A295},
  year={2006},
  publisher={IOP Publishing}
}

@article{koerver2017capacity,
  title={Capacity fade in solid-state batteries: interphase formation and chemomechanical processes in nickel-rich layered oxide cathodes and lithium thiophosphate solid electrolytes},
  author={Koerver, Raimund and Ayg{\"u}n, Isabel and Leichtwei{\ss}, Thomas and Dietrich, Christian and Zhang, Wenbo and Binder, Jan O and Hartmann, Pascal and Zeier, Wolfgang G and Janek, J{\"u}rgen},
  journal={Chemistry of Materials},
  volume={29},
  number={13},
  pages={5574--5582},
  year={2017},
  publisher={ACS Publications}
}

@article{bucci2018mechanical,
  title={Mechanical instability of electrode-electrolyte interfaces in solid-state batteries},
  author={Bucci, Giovanna and Talamini, Brandon and Balakrishna, Ananya Renuka and Chiang, Yet-Ming and Carter, W Craig},
  journal={Physical Review Materials},
  volume={2},
  number={10},
  pages={105407},
  year={2018},
  publisher={APS}
}

@article{xiang2017accommodating,
  title={Accommodating high transformation strains in battery electrodes via the formation of nanoscale intermediate phases: operando investigation of olivine $\mathrm{NaFePO_4}$},
  author={Xiang, Kai and Xing, Wenting and Ravnsb{\ae}k, Dorthe B and Hong, Liang and Tang, Ming and Li, Zheng and Wiaderek, Kamila M and Borkiewicz, Olaf J and Chapman, Karena W and Chupas, Peter J and others},
  journal={Nano letters},
  volume={17},
  number={3},
  pages={1696--1702},
  year={2017},
  publisher={ACS Publications}
}

@article{chu1995analysis,
  title={Analysis of microstructures in $\mathrm{Cu}$-14.0\%$\mathrm{Al}$-3.9\%$\mathrm{Ni}$ by energy minimization},
  author={Chu, C and James, RD},
  journal={Le Journal de Physique IV},
  volume={5},
  number={C8},
  pages={C8--143},
  year={1995},
  publisher={EDP sciences}
}

@article{zhang2022DFT,
  title={},
  author={Zhang, D and Renuka Balakrishna, A},
  journal={Manuscript in preparation},
  volume={},
  number={},
  pages={},
  year={2022},
  publisher={}
}

@article{ericksen2008cauchy,
  title={On the cauchy-born rule},
  author={Ericksen, J. L.},
  journal={Mathematics and mechanics of solids},
  volume={13},
  number={3-4},
  pages={199--220},
  year={2008},
  publisher={Sage Publications Sage UK: London, England}
}

@incollection{james2005way,
  title={A way to search for multiferroic materials with ``unlikely'' combinations of physical properties},
  author={James, RD and Zhang, Zhiyong},
  booktitle={Magnetism and structure in functional materials},
  pages={159--175},
  year={2005},
  publisher={Springer}
}

@article{kang2006factors,
  title={Factors that affect $\mathrm{Li}$ mobility in layered lithium transition metal oxides},
  author={Kang, Kisuk and Ceder, Gerbrand},
  journal={Physical Review B},
  volume={74},
  number={9},
  pages={094105},
  year={2006},
  publisher={APS}
}

@article{james1998magnetostriction,
  title={Magnetostriction of martensite},
  author={James, Richard D and Wuttig, Manfred},
  journal={Philosophical magazine A},
  volume={77},
  number={5},
  pages={1273--1299},
  year={1998},
  publisher={Taylor \& Francis}
}

@article{li2005domain,
  title={Domain switching in polycrystalline ferroelectric ceramics},
  author={Li, JY and Rogan, RC and {\"U}st{\"u}ndag, E and Bhattacharya, K},
  journal={Nature materials},
  volume={4},
  number={10},
  pages={776--781},
  year={2005},
  publisher={Nature Publishing Group}
}

@article{naumov2015mechanically,
  title={Mechanically responsive molecular crystals},
  author={Naumov, Pance and Chizhik, Stanislav and Panda, Manas K and Nath, Naba K and Boldyreva, Elena},
  journal={Chemical reviews},
  volume={115},
  number={22},
  pages={12440--12490},
  year={2015},
  publisher={ACS Publications}
}

@article{chluba2015ultralow,
  title={Ultralow-fatigue shape memory alloy films},
  author={Chluba, Christoph and Ge, Wenwei and Lima de Miranda, Rodrigo and Strobel, Julian and Kienle, Lorenz and Quandt, Eckhard and Wuttig, Manfred},
  journal={Science},
  volume={348},
  number={6238},
  pages={1004--1007},
  year={2015},
  publisher={American Association for the Advancement of Science}
}

@article{chen2013study,
  title={Study of the cofactor conditions: conditions of supercompatibility between phases},
  author={Chen, Xian and Srivastava, Vijay and Dabade, Vivekanand and James, Richard D},
  journal={Journal of the Mechanics and Physics of Solids},
  volume={61},
  number={12},
  pages={2566--2587},
  year={2013},
  publisher={Elsevier}
}

@article{islam2014lithium,
  title={Lithium and sodium battery cathode materials: computational insights into voltage, diffusion and nanostructural properties},
  author={Islam, M Saiful and Fisher, Craig AJ},
  journal={Chemical Society Reviews},
  volume={43},
  number={1},
  pages={185--204},
  year={2014},
  publisher={Royal Society of Chemistry}
}

@article{urban2016computational,
  title={Computational understanding of $\textrm{Li}$-ion batteries},
  author={Urban, Alexander and Seo, Dong-Hwa and Ceder, Gerbrand},
  journal={npj Computational Materials},
  volume={2},
  number={1},
  pages={1--13},
  year={2016},
  publisher={Nature Publishing Group}
}

@article{parker2022alloy,
  title={Alloying-Induced Pre-Transformation as a Means of Accessing Extended Solid-Solution Regimes in an Intercalation Cathode},
  author={Schofield, Parker and Luo, Yuting and Zhang, Delin and Santos, David and Zaheer, Wasif and Agbeworvi, George and Handy, Joseph and Andrews, Justin and Braham, Erick and Renuka Balakrishna, Ananya and Banerjee, Sarbajit},
  journal={Manuscript in preparation},
  year={2022},
}

@article{martinolich2020controlling,
  title={Controlling covalency and anion redox potentials through anion substitution in Li-rich chalcogenides},
  author={Martinolich, Andrew J and Zak, Joshua J and Agyeman-Budu, David N and Kim, Seong Shik and Bashian, Nicholas H and Irshad, Ahamed and Narayan, Sri R and Melot, Brent C and Nelson Weker, Johanna and See, Kimberly A},
  journal={Chemistry of Materials},
  volume={33},
  number={1},
  pages={378--391},
  year={2020},
  publisher={ACS Publications}
}

@article{thomas2021comparing,
  title={Comparing crystal structures with symmetry and geometry},
  author={Thomas, John C and Natarajan, Anirudh Raju and Van der Ven, Anton},
  journal={npj Computational Materials},
  volume={7},
  number={1},
  pages={1--11},
  year={2021},
  publisher={Nature Publishing Group}
}

@article{xiao2019computational,
  title={Computational screening of cathode coatings for solid-state batteries},
  author={Xiao, Yihan and Miara, Lincoln J and Wang, Yan and Ceder, Gerbrand},
  journal={Joule},
  volume={3},
  number={5},
  pages={1252--1275},
  year={2019},
  publisher={Elsevier}
}

@article{jun2022lithium,
  title={Lithium superionic conductors with corner-sharing frameworks},
  author={Jun, KyuJung and Sun, Yingzhi and Xiao, Yihan and Zeng, Yan and Kim, Ryounghee and Kim, Haegyeom and Miara, Lincoln J and Im, Dongmin and Wang, Yan and Ceder, Gerbrand},
  journal={Nature Materials},
  pages={1--8},
  year={2022},
  publisher={Nature Publishing Group}
}

@article{chen2014pyrophosphate,
  title={Pyrophosphate $\mathrm{Na_2FeP_2O_7}$ as a low-cost and high-performance positive electrode material for sodium secondary batteries utilizing an inorganic ionic liquid},
  author={Chen, Chih-Yao and Matsumoto, Kazuhiko and Nohira, Toshiyuki and Hagiwara, Rika and Orikasa, Yuki and Uchimoto, Yoshiharu},
  journal={Journal of Power Sources},
  volume={246},
  pages={783--787},
  year={2014},
  publisher={Elsevier}
}

@article{kim2015anomalous,
  title={Anomalous Jahn--Teller behavior in a manganese-based mixed-phosphate cathode for sodium ion batteries},
  author={Kim, Hyungsub and Yoon, Gabin and Park, Inchul and Park, Kyu-Young and Lee, Byungju and Kim, Jongsoon and Park, Young-Uk and Jung, Sung-Kyun and Lim, Hee-Dae and Ahn, Docheon and others},
  journal={Energy \& Environmental Science},
  volume={8},
  number={11},
  pages={3325--3335},
  year={2015},
  publisher={Royal Society of Chemistry}
}

@article{patoux2002lithium,
  title={Lithium insertion into titanium phosphates, silicates, and sulfates},
  author={Patoux, S{\'e}bastien and Masquelier, Christian},
  journal={Chemistry of Materials},
  volume={14},
  number={12},
  pages={5057--5068},
  year={2002},
  publisher={ACS Publications}
}

@article{thackeray1997manganese,
  title={Manganese oxides for lithium batteries},
  author={Thackeray, Michael M},
  journal={Progress in Solid State Chemistry},
  volume={25},
  number={1-2},
  pages={1--71},
  year={1997},
  publisher={Elsevier}
}

@article{yin2020structural,
  title={Structural evolution at the oxidative and reductive limits in the first electrochemical cycle of $\mathrm{Li_{1.2}Ni_{0.13}Mn_{0.54}Co_{0.13}O_2}$},
  author={Yin, Wei and Grimaud, Alexis and Rousse, Gwena{\"e}lle and Abakumov, Artem M and Senyshyn, Anatoliy and Zhang, Leiting and Trabesinger, Sigita and Iadecola, Antonella and Foix, Dominique and Giaume, Domitille and others},
  journal={Nature Communications},
  volume={11},
  number={1},
  pages={1--11},
  year={2020},
  publisher={Nature Publishing Group}
}

@article{zhuo2006preparation,
  title={The preparation of $\mathrm{NaV_\mathit{1-x}Cr_\mathit{x}PO_4F}$ cathode materials for sodium-ion battery},
  author={Zhuo, Haitao and Wang, Xianyou and Tang, Anping and Liu, Zhiming and Gamboa, Sergio and Sebastian, PJ},
  journal={Journal of power sources},
  volume={160},
  number={1},
  pages={698--703},
  year={2006},
  publisher={Elsevier}
}


\section*{Supplementary material}

# Supplementary Information
(A systematic search and microstructural analysis of intercalation electrodes)

Delin Zhang[1] and Ananya Renuka Balakrishna[1*]

[1]Aerospace and Mechanical Engineering, University of Southern California, Los Angeles, CA 90089
[*]Corresponding author, Email: renukaba@usc.edu

In these notes, we briefly describe the symbols and terminologies used in the paper, provide further details on the structural transformation algorithm, and outline the crystallographic design rules necessary to form shape-memory-like microstructures. We explain these concepts using representative intercalation compounds such as $Li_xFe_2(MoO_4)_3$ and $Li_xMn_2O_4$.

# Symbols and Terminologies

| Notation | Brief description |
| --- | --- |
| $e_i$ | Lattice vector |
| $\mathbf{E} = \{\mathbf{e}_1, \mathbf{e}_2, \mathbf{e}_3\}$ | Primitive unit cell spanned by $e_i (i = 1, 2, 3)$ |
| $\mathbf{E}_R$, $\mathbf{E}_I$ | Primitive unit cell of reference and intercalated phase |
| $\mathbf{P}_R$, $\mathbf{P}_I$ | Lattice correspondence matrix of reference and intercalated phase |
| $\mathbf{E}^A = \mathbf{E}\mathbf{P}^A$ | Conventional cell related to $\mathbf{E}$ via a lattice correspondence $\mathbf{P}^A$ |
| $n$ | Number of intercalation compounds |
| $N$ | Number of variants |
| $\mathbf{I}$ | Identity matrix |
| $\mathbf{F}$ | Deformation gradient tensor |
| $\mathbf{U}$, $\mathbf{U}_N$ | Transformation stretch tensor. Variants $\mathbf{U}_N$ are related to each other via a symmetry operation |
| $\mathbf{M}$ | Necessary and sufficient condition for a monoclinic compound to form self-accommodating microstructure (see Ref. [1] Section 3.4) |
| $\mathbf{Q}$, $\mathbf{Q}'$ | Rotation matrices |
| $\mathbf{a}$, $\mathbf{b}$, $\hat{\mathbf{m}}$, $\hat{\mathbf{n}}$ | Vectors |
| $K$ | Twin plane direction |
| $k$ | A constant which can take a value of $\pm 1$ |
| $\lambda_i$ | Eigenvalues when solving $\mathbf{QU}_I - \mathbf{I} = \mathbf{b} \otimes \hat{\mathbf{m}}$ |
| $f$ | Volume fraction |
| $\theta$ | Orientation of the twin plane |
| $|\det \mathbf{U} - 1| = 0$ | Volume preservation |
| $|\lambda_2 - 1| = 0$ | Condition to form stress-free interface |
| $p$, $q$ | Number of host-atoms in reference and intercalated unit cells |
| $m = \frac{p}{q}$ | Ratio of host-atoms |
| $d$ | largest edge ratio |
| $\{a, b, c, \alpha, \beta, \gamma\}$ | Lattice parameters |

Supplementary Table 1: Summary of symbols

| Terminology | Description |
| --- | --- |
| Deformation gradient $\mathbf{F}$ | The deformation gradient $\mathbf{F}$ quantifies the deformation in continuum mechanics. It is a second order tensor which maps line elements in the reference configuration into line elements (consisting of the same material particles) in the deformed configuration. |
| Stretch tensor $\mathbf{U}$ | During phase transformation, the reference phase typically has a different lattice geometry from the transformed (or intercalated) phase. A stretch tensor maps a unit cells of the reference phase to the corresponding unit cell of the transformed phase. This tensor describes the local stretching (or contraction) at a material point. |
| Eigenvalues | A set of scalars associated with a linear system of equations. In our calculations, these values correspond to the lattice stretches during transformation. |
| Variant | Variants are lattices of different orientations belonging to the same phase of the material. These variants are related to each other via a symmetry operation. |
| Bravais lattice | A Bravais lattice is an infinite set of points in three-dimensional space generated by the translation of a single point $\mathbf{o}$ through three linearly independent lattice vectors $\mathbf{E} = \{\mathbf{e}_1, \mathbf{e}_2, \mathbf{e}_3\}$. |
| Primitive unit cell | The lattice vectors $\mathbf{e}_i$ enclosing the smallest repeatable volume constitute a primitive unit cell $\mathbf{E}$. |
| Conventional unit cell | The latter lattice vectors $\mathbf{e}'_i$ enclosing a larger volume (relative to $\mathbf{E}$) constitute a conventional unit cell $\mathbf{E}'$. A conventional cell of a crystal structure can be described in multiple ways, however, all the conventional and the primitive cells generate the same set of lattice points in 3D space. |
| Lattice correspondence $\mathbf{P}$ | A conventional cell is related to the primitive cell via a lattice correspondence matrix $\mathbf{P}$, i.e., $\mathbf{E}' = \mathbf{E}\mathbf{P}$ |
| Frobenius norm $\|.\|$ | A matrix norm which is defined as the square root of the sum of the absolute squares of elements in a matrix. It is also equal to the square root of the matrix trace of $\mathbf{A}^{\mathrm{T}}\mathbf{A}$, i.e., $\|\mathbf{A}\| = \sqrt{\mathrm{Tr}(\mathbf{A}^{\mathrm{T}}\mathbf{A})}$ |

Supplementary Table 2: Definition of terminology

# Structural transformation

In this section, we describe how we compute individual lattice deformations from the crystal structure database in detail. Following Chen et al. [2], we determine the transformation pathway between reference and intercalated lattices by minimizing a distance function. This minimization technique helps to identify a stretch tensor–with minimum structural distortion–from a group of potential stretch tensors that map reference to intercalated phases. Below, we outline the three tasks to derive an optimal deformation tensor, see Supplementary Figure 1.

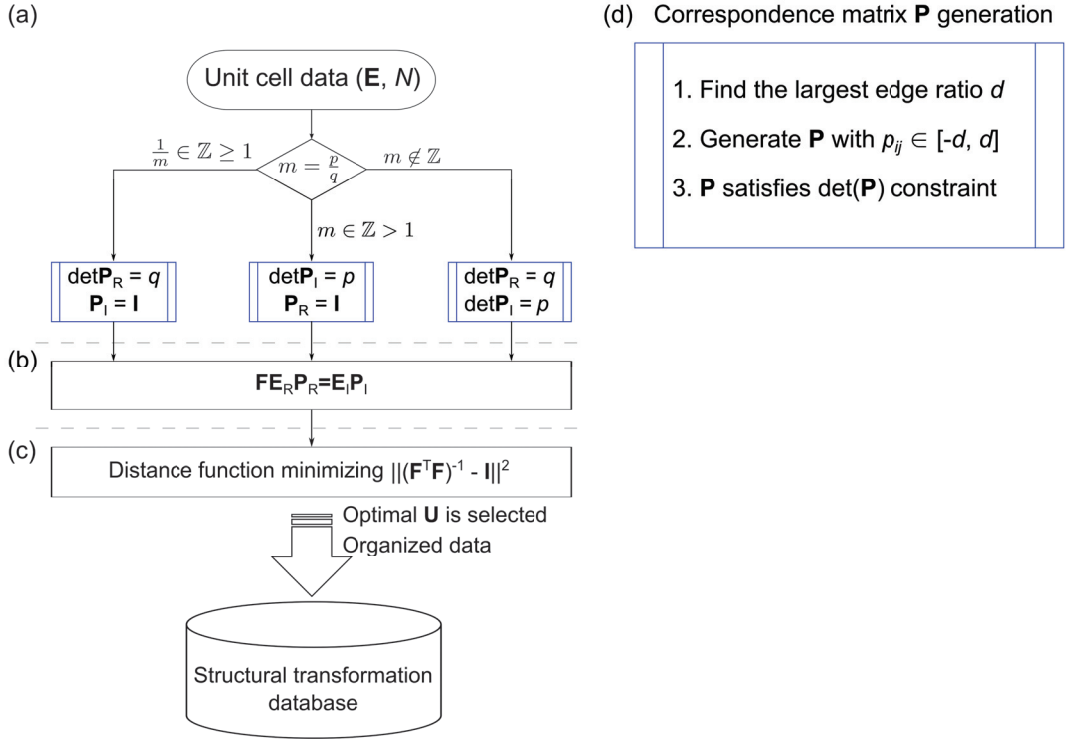

Supplementary Figure 1: The flowchart illustrates the three steps of our algorithm. (a) We determine the primitive basis $\mathbf{E}_R, \mathbf{E}_I$ and correspondence matrices $\mathbf{P}_R, \mathbf{P}_I$ of the reference and intercalated phases. The conventional unit cells are represented in the form of primitive unit cells, $\mathbf{E}' = \mathbf{EP}$, and the irreducible fraction $m$ is the ratio of atoms in the reference unit cell to the intercalated unit cell. (b) We next determine all potential structural transformation tensors $\mathbf{F}$ that map a unit cell of the reference phase to that of the intercalated phase. (c) We minimize the distance function to determine the optimal stretch tensor $\mathbf{U} = \sqrt{\mathbf{F}^T\mathbf{F}}$ and its variants. The computed stretch tensors are stored in the structural transformation database and used for microstructural analysis. (d) We identify all lattice correspondence matrices $\mathbf{P}$ in (a) according to the largest edge ratio $d$ for the reference and intercalated primitive cells of a given compound.

## 1. Identify unit cells

In the first task, we determine the primitive bases and the lattice correspondence matrices for the reference and intercalated phases of a given compound, see Supplementary Figure 1(a). The primi-

4

tive cell is the smallest, repeatable unit that generates the crystal structure. However, the primitive cells of the reference and intercalated phases for a given compound are not always described to contain the same number of host-atoms.[1] For example, in Li$_x$Fe$_2$(MoO$_4$)$_3$ intercalation compound, the primitive lattice of the reference phase contains twice as many atoms as the primitive lattice of the intercalated phase, see Supplementary Table 3 and Supplementary Figure 2. Consequently, the intercalated phase should be transformed to a larger unit cell (via a suitable correspondence matrix) to determine the structural transformation pathway. In this task, we identify all potential correspondence matrices $\mathbf{P}$ for the reference (and/or intercalated) primitive phases. Below, we outline the specific steps to compute $\mathbf{P}$ using a pair of primitive unit cells as an example. Similar steps are followed to compute $\mathbf{P}$ of the primitive cell for one phase to match the fixed conventional cell for the other.

(i) We calculate the ratio of host-atoms in the unit cells of the reference and intercalated phases for a given compound, an irreducible fraction $m = \frac{p}{q}$ with $p, q \in \mathbb{Z}$. This ratio can be of three types $m \in \mathbb{Z}$, $\frac{1}{m} \in \mathbb{Z}$, and $m = \notin \mathbb{Z}$. For example, the ratio of atoms in the reference and intercalated phases of Li$_x$Fe$_2$(MoO$_4$)$_3$ compound (see Supplementary Table 3) is $m = 2$. We next use the values of $m, p, q$, to identify all potential correspondence matrices in steps (ii) – (iii).

(ii) We determine the largest edge ratio $d$, which is rounded-off to the nearest integer, for the reference and intercalated primitive cells of a given compound. For example, in Li$_x$Fe$_2$(MoO$_4$)$_3$ compound the largest edge corresponds to the reference phase (e.g., $a_\mathrm{R} = 15.7380$Å) and the smallest edge corresponds to the intercalated phase (e.g., $b_\mathrm{I} = 9.4837$Å). The edge ratio is therefore $d = \frac{15.7380\text{Å}}{9.4837\text{Å}} \approx 2$. Physically, this ratio determines the bounds on the elements $p_{ij}$ of the correspondence matrix $\mathbf{P}$.

(iii) We next construct lattice correspondence matrices $\mathbf{P}$ such that each element satisfies $p_{ij} \in [-d, d]$. A total of $d^9$ matrices can be constructed, however, we only store matrices that satisfy $\det(\mathbf{P}_\mathrm{R}) = q$ and $\det(\mathbf{P}_\mathrm{I}) = p$ in the structural database.[2] For Li$_x$Fe$_2$(MoO$_4$)$_3$ compound a total of 147,696 correspondence matrices satisfied $\det(\mathbf{P}_\mathrm{I}) = m = 2$.

We follow steps (i)–(iii) to determine the primitive bases and suitable correspondence matrices for all intercalation compounds and use them to determine the stretch tensors $\mathbf{U}$.

---

[1] Please note that in intercalation materials, the total number of atoms refers to the atoms of the host-material and does not account for the guest (or intercalating) species.

[2] This is to ensure the volume change between the reference and intercalated phases matches the volume scaling of the correspondence matrix. However, when $p = 1$ or $q = 1$, we define $\mathbf{P}_\mathrm{R} = \mathbf{I}$ or $\mathbf{P}_\mathrm{I} = \mathbf{I}$, respectively. For further details please see Ref. [2].

## 2. Compute stretch tensor

In the second task, we compute the transformation stretch tensor $\mathbf{U}$ that maps a unit cell of the reference phase to the corresponding unit cell of the intercalated phase. The general form for the mapping is given by:

$$\mathbf{F}\mathbf{E}_\mathrm{R}\mathbf{P}_\mathrm{R} = \mathbf{E}_\mathrm{I}\mathbf{P}_\mathrm{I} \tag{S.1}$$

in which $\mathbf{F}$ is the deformation gradient, and the transformation stretch tensor $\mathbf{U}$ is the unique positive-definite square root of $\mathbf{F}^\mathrm{T}\mathbf{F}$. This stretch tensor $\mathbf{U}$ is unique for a given pair of correspondence matrices $\mathbf{P}_\mathrm{R}, \mathbf{P}_\mathrm{I}$, and we determine all possible stretch tensors by cycling through the set of correspondence matrices. For example, in $\mathrm{Li}_x\mathrm{Fe}_2(\mathrm{MoO}_4)_3$ we determine a total of 36,924 stretch tensors by cycling through its set of correspondence matrices.

## 3. Minimize the distance function

In the third task, we determine the optimal lattice correspondence (and thus the transformation stretch tensor) that minimizes the total strain of the structural transformation. We note that multiple lattice correspondences and transformation stretch tensors could describe different structural transformation pathways between the reference and intercalated phases. Following Chen et al. [2], we introduce a distance function that represents the total strain of a structural transformation:

$$\mathrm{dist}(\mathbf{P}_\mathrm{R}, \mathbf{P}_\mathrm{I}, \mathbf{E}_\mathrm{R}, \mathbf{E}_\mathrm{I}) = \left\|\mathbf{U}^{-2} - \mathbf{I}\right\|^2 \tag{S.2}$$

in which $\|.\|$ is the Frobenius norm, $\|\mathbf{A}\|^2 = \mathbf{A} : \mathbf{A} = \mathrm{Tr}(\mathbf{A}^\mathrm{T}\mathbf{A})$. By minimizing Eq. (S.2) over the set of stretch tensors, we determine the optimal lattice correspondence and the transformation stretch tensor for a given intercalation compound. In $\mathrm{Li}_x\mathrm{Fe}_2(\mathrm{MoO}_4)_3$, we identify a stretch tensor $\mathbf{U}$ that minimizes the distance function, see Supplementary Table 3. In principle, this algorithm can be applied to determine stretch tensors for any crystalline material undergoing a first-order phase transformation, however, for the present work we apply our theoretical framework to intercalation compounds commonly used in batteries. We use the stretch tensor identified by our algorithm to investigate whether the material can form shape-memory-like microstructures.

|  | Fe$_2$(MoO$_4$)$_3$ | Li$_2$Fe$_2$(MoO$_4$)$_3$ |
|---|---|---|
| Space group | $P2_1/c$ | $Pbcn$ |
| Crystal symmetry | monoclinic | orthorhombic |
| Volume(Å$^3$) | 2156.5 | 1141.98 |
| $\{a,b,c\}$(Å) | $\{15.7380, 9.2352, 15.7040\}$ | $\{12.8946, 9.4837, 9.3384\}$ |
| $\{\alpha,\beta,\gamma\}$(°) | $\{90, 109.1209, 90\}$ | $\{90, 90, 90\}$ |
| Atoms and their Wyckoff position | Mo1 − Mo6 : 4e | Mo1 : 4c  Mo2 : 8d |
| Predicted stretch tensor **U** | $\begin{bmatrix} 1.0116 & 0 & 0.0080 \\ 0 & 1.0269 & 0 \\ 0.0080 & 0 & 1.0196 \end{bmatrix}$ | |

Supplementary Table 3: Detailed information of Fe$_2$(MoO$_4$)$_3$ and Li$_2$Fe$_2$(MoO$_4$)$_3$ [3].

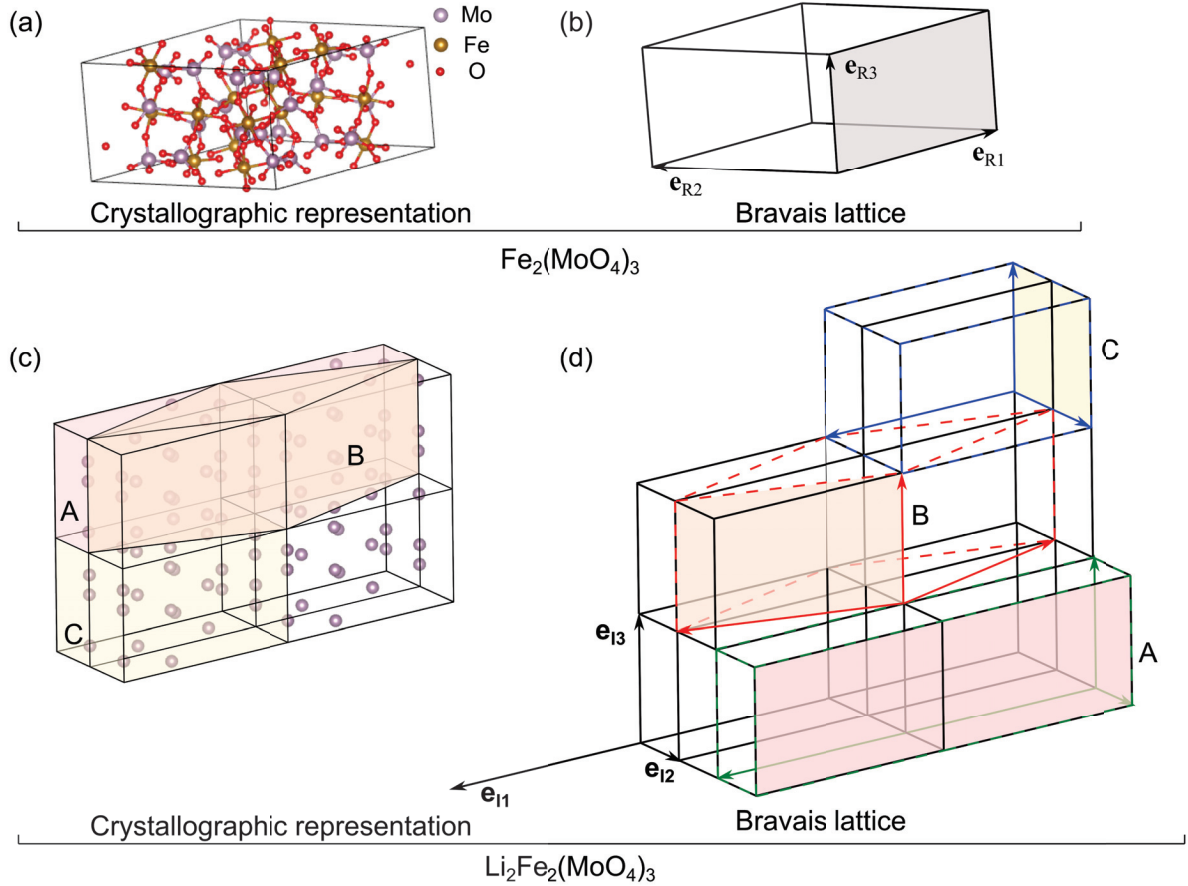

Supplementary Figure 2: Crystallographic and Bravais lattice representations of unit cells of (a-b) $Fe_2(MoO_4)_3$ (reference phase) and (c-d) $Li_2Fe_2(MoO_4)_3$ (intercalated phase). (a) and (b) The unit cell $Fe_2(MoO_4)_3$ spanned by vectors $\mathbf{e}_{R1}, \mathbf{e}_{R2}, \mathbf{e}_{R3}\}$ describe the primitive unit cell of $Fe_2(MoO_4)_3$. (c) and (d) Similarly, the vectors $\{\mathbf{e}_{I1}, \mathbf{e}_{I2}, \mathbf{e}_{I3}\}$ enclose the primitive unit cell of the intercalated phase $Li_2Fe_2(MoO_4)_3$. (c-d) also show three representative examples of conventional unit cells identified by 'A', 'B' and 'C'. On minimizing the distance function across all potential conventional cells we identify that $Fe_2(MoO_4)_3$ transforming to the B-type $Li_2Fe_2(MoO_4)_3$ unit cell has minimum structural distortions. This is consistent with the experimental results [3].

# Crystallographic Theory of Martensites

In this section, we give an overview of lattice deformation and outline the necessary conditions to form shape-memory-like microstructures (i.e., twin boundaries, austenite-martensite interface, self-accommodating microstructures, and $\lambda_2 = 1$ microstructures). These microstructures form in materials with very specific lattice geometries, which satisfy the crystallographic design rules described below. In our analysis, we use these design rules as a quantitative guide to investigate whether any known intercalation compound can form shape-memory-like microstructures.

## Twin interface

Twin interfaces are energy minimization deformations in which lattices rotate and/or shear to fit compatibly with each other. These twins form in materials that undergo lattice-symmetry lowering during phase transformation.

For example, in Supplementary Figure 3, a cubic lattice in the reference phase transforms into a lower-symmetry tetragonal lattice. This transformation can generate three variants of the tetragonal lattices, which are described by stretch tensors $\mathbf{U}_1, \mathbf{U}_2, \mathbf{U}_3$.[3] A twin boundary forms when any two lattice variants, with deformation tensors $\mathbf{U}_I$ and $\mathbf{U}_J$, satisfy the kinematic compatibility condition:

$$\mathbf{Q}\mathbf{U}_I - \mathbf{U}_J = \mathbf{a} \otimes \hat{\mathbf{n}} \tag{S.3}$$

for a given rotation matrix $\mathbf{Q}$ and vectors $\mathbf{a} \neq 0$ and $\hat{\mathbf{n}}$. The deformation tensors have positive determinants. For our purposes, we note that Eq. (S.3) has a solution if and only if its eigenvalues $\lambda_i$, which describe relative lattice stretches, satisfy $\lambda_1 \leq 1$, $\lambda_2 = 1$ and $\lambda_3 \geq 1$. The solution to Eq. (S.3) then describes a twin plane that connects the two lattice variants $\mathbf{U}_I$ and $\mathbf{U}_J$ coherently. These twin boundaries often have zero (or negligible) elastic energy, and thus stabilizing these microstructural features would reduce internal stresses in intercalation materials.

Let us construct a twin interface using $Li_2Mn_2O_4$ as a representative example. Supplementary Table 4 lists the stretch tensors for $Li_2Mn_2O_4$, and we construct a twin interface formed between two variants $\mathbf{U}_1, \mathbf{U}_2$. These variants form a twin interface if they satisfy the compatibility condition Eq. (S.3) for a given rotation $\mathbf{Q}$ and vectors $\mathbf{a} \neq 0$ and $\hat{\mathbf{n}}$. Following Ref. [4] we construct the twin interface as follows:

1. Calculate the matrix $\mathbf{C} = (\mathbf{U}_1 \mathbf{U}_2^{-1})^{\mathrm{T}} \mathbf{U}_1 \mathbf{U}_2^{-1}$

2. If $\mathbf{C} = \mathbf{I}$, then there is no solution to Eq. (S.3)

---

[3] The number of variants $N$ is the ratio of the number of rotations in the point groups of the reference and intercalated phases, respectively [4]. For example, in the case of the cubic to tetragonal transformation, the number of rotations in the point groups of the cubic to tetragonal phases are 24 and 8, respectively, and the number of variants is $N = 24/8 = 3$.

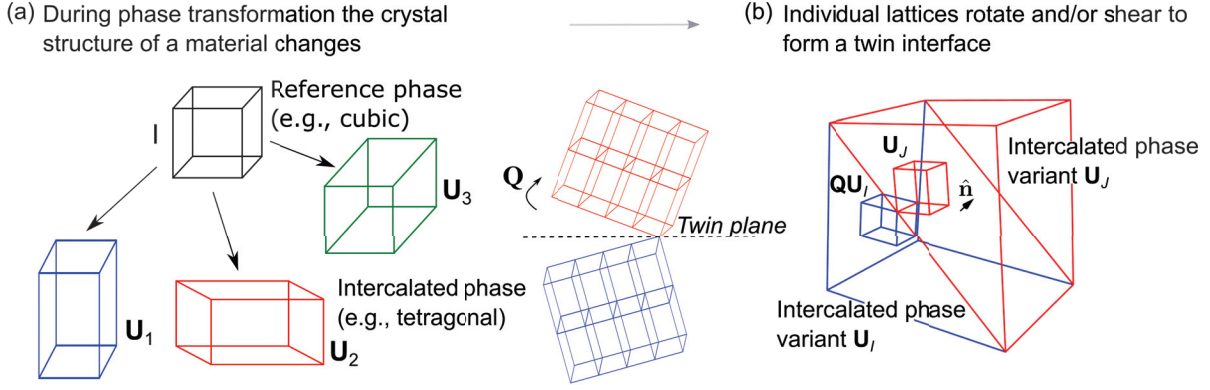

Supplementary Figure 3: A schematic illustration of the twin interface formation. (a) During phase transformation the crystal structure of the material changes. For example, a high-symmetry cubic lattice $\mathbf{I}$ (reference phase) transforms to low-symmetry tetragonal lattices (intercalated) $\mathbf{U}_N$. This symmetry-lowering transformation generates three distinct variants $\mathbf{U}_1, \mathbf{U}_2, \mathbf{U}_3$. (b) These lattice variants then rotate and/or shear to form a twin interface. The twin interface in subfigure (b) is generated by solving Eq. (S.3) for a spinel compound [5]

3. If $\mathbf{C} \neq \mathbf{I}$, it is possible to verify that $\mathbf{C}$ is symmetric and positive-definite. Thus the matrix $\mathbf{C}$ can be given by:

$$\mathbf{C} = \lambda_1 \hat{\mathbf{e}}_1 \otimes \hat{\mathbf{e}}_1 + \lambda_2 \hat{\mathbf{e}}_2 \otimes \hat{\mathbf{e}}_2 + \lambda_3 \hat{\mathbf{e}}_3 \otimes \hat{\mathbf{e}}_3 \tag{S.4}$$

where $\lambda_i > 0$ are eigenvalues of the matrix $\mathbf{C}$, and $\hat{\mathbf{e}}_i$ are eigenvectors of $\mathbf{C}$ corresponding to $\lambda_i$. Eq. (S.3) has a solution if and only if the eigenvalues satisfy $\lambda_1 \leq \lambda_2 = 1 \leq \lambda_3$.

4. If above-mentioned conditions are satisfied, then there are exactly two solutions give by:

$$\mathbf{a} = \rho \left( \sqrt{\frac{\lambda_3(1-\lambda_1)}{\lambda_3 - \lambda_1}} \hat{\mathbf{e}}_1 + k\sqrt{\frac{\lambda_1(\lambda_3 - 1)}{\lambda_3 - \lambda_1}} \hat{\mathbf{e}}_3 \right)$$

$$\hat{\mathbf{n}} = \frac{\sqrt{\lambda_3} - \sqrt{\lambda_1}}{\rho\sqrt{\lambda_3 - \lambda_1}} \left( -\sqrt{1-\lambda_1} \mathbf{U}_2^T \hat{\mathbf{e}}_1 + k\sqrt{\lambda_3 - 1} \mathbf{U}_2^T \hat{\mathbf{e}}_3 \right) \tag{S.5}$$

where $k = \pm 1$, $\rho \neq 0$ is chosen to make $|\hat{\mathbf{n}}| = 1$. Choosing $k = 1$ gives one solution while $k = -1$ gives the other. In both situation, we obtain $\mathbf{Q}$ by substituting $\mathbf{a}$ and $\hat{\mathbf{n}}$ back into Eq. (S.3).

5. We substitute the values of stretch tensors $\mathbf{U}_1, \mathbf{U}_2$ for $Li_2Mn_2O_4$ and list its twin solutions in Supplementary Table 4. We follow similar steps to compute the twin interfaces for other intercalation compounds listed in the main paper.

| Stretch tensor | | Twin solutions | |
|---|---|---|---|
| $\mathbf{U}_1 = \begin{bmatrix} 0.9690 & 0 & 0 \\ 0 & 1.1226 & 0 \\ 0 & 0 & 0.9690 \end{bmatrix}$ | | $\mathbf{a}_1 = [0.2319, 0.2001, 0]$ <br> $\hat{\mathbf{n}}_1 = [-0.7071, 0.7071, 0]$ <br> $K_1 = (-0.6534, 0.7570, 0)$ | $\mathbf{Q}_1 = \begin{bmatrix} 0.9893 & 0.1461 & 0 \\ -0.1461 & 0.9893 & 0 \\ 0 & 0 & 1 \end{bmatrix}$ |
| $\mathbf{U}_2 = \begin{bmatrix} 1.1226 & 0 & 0 \\ 0 & 0.9690 & 0 \\ 0 & 0 & 0.9690 \end{bmatrix}$ | | $\mathbf{a}_2 = [0.2319, -0.2001, 0]$ <br> $\hat{\mathbf{n}}_2 = [-0.7071, -0.7071, 0]$ <br> $K_2 = (-0.6534, -0.7570, 0)$ | $\mathbf{Q}_2 = \begin{bmatrix} 0.9893 & -0.1461 & 0 \\ 0.1461 & 0.9893 & 0 \\ 0 & 0 & 1 \end{bmatrix}$ |
| $\mathbf{U}_3 = \begin{bmatrix} 0.9690 & 0 & 0 \\ 0 & 0.9690 & 0 \\ 0 & 0 & 1.1226 \end{bmatrix}$ | | | |

Supplementary Table 4: Twin solutions for $\mathbf{U}_1$ and $\mathbf{U}_2$ variants of $Li_2Mn_2O_4$

## Austenite-Martensite interface

In phase transformation materials, the twin interfaces are rarely found in isolation. Instead, complex microstructures, such as those in Supplementary Figure 4(a), that comprises a mixture of finely twinned interfaces are commonly observed. These complex microstructures form as a consequence of energy minimization and are commonly observed in ferroelastic materials, such as the shape memory alloys, and have recently been reported in intercalation cathodes.

Ball and James [6] explain the origin of the austenite-martensite microstructures in shape-memory alloys using an energy minimization theory. In this crystallographic theory, the free energy of the material is described as a function of the lattice deformation gradient that penalizes the elastic energy arising from lattice misfit. Minimizing this energy then results in the formation of finely twinned microstructures. For example, during the cubic-to-tetragonal phase transformation, the phase boundary separating the two phases is elastically stressed because of the lattice misfit strains, see Supplementary Figure 4(a). This elastic energy can be minimized by forming a finely twinned mixture in the tetragonal phase that fits coherently with the cubic phase.

Analytically, an austenite-martensite microstructure forms when any two lattice variants, with deformation tensors $\mathbf{U}_I$ and $\mathbf{U}_J$, satisfy the kinematic compatibility condition for some scalar $0 \leq f \leq 1$:[4]

$$\mathbf{Q}'(f\mathbf{Q}\mathbf{U}_J + (1-f)\mathbf{U}_I) = \mathbf{I} + \mathbf{b} \otimes \hat{\mathbf{m}}. \tag{S.6}$$

---

[4]The scalar $f$ corresponds to the volume fraction of different variants in the martensite mixture.

Here the reference lattice is represented by $\mathbf{I}$, rotation matrices by $\mathbf{Q}, \mathbf{Q}'$, and vectors $\mathbf{b} \neq 0$ and $\hat{\mathbf{m}}$.

The austenite-martensite interface describes energy minimizing deformations that reduce coherency stresses at the phase boundary. This interface forms for specific lattice geometry conditions and in materials that can undergo a symmetry-lowering phase transformation. These microstructures, if stabilized in intercalation electrodes, would address structural problems related to stressed phase boundaries in cathodes and help mitigate the chemo-mechanical challenges during the charging/discharging process.

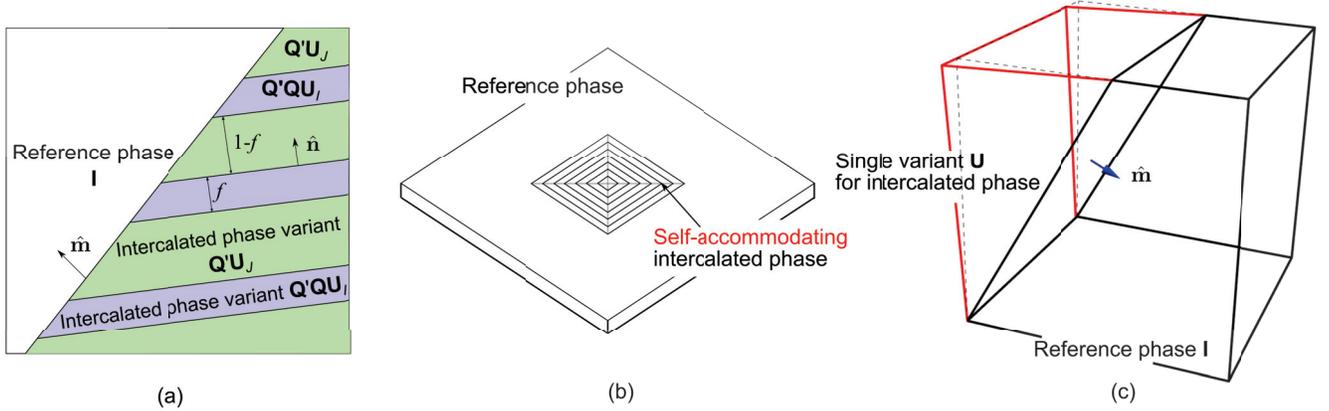

Supplementary Figure 4: (a) Schematic illustration of an austenite-martensite interface. Here, a finely twinned microstructure (with variants $\mathbf{U}_I, \mathbf{U}_J$) forms a coherent interface with the reference phase. (b) In a self-accommodating microstructure, finely twinned variants coherently arrange themselves to occupy a region in the reference phase, without any macroscopic change in shape. (c) In the $\lambda_2 = 1$ microstructure, a perfectly compatible interface forms between the reference phase and a single variant of the intercalated phase. This interface is stress-free.

As an example, we construct austenite/martensite interfaces for $Li_2Mn_2O_4$ using its stretch tensors from Supplementary Table 4. Following Ref. [4], we identify rotation tensors $\mathbf{Q}'$, vectors $\mathbf{b}, \hat{\mathbf{m}}$, volume fraction of the martensite mixture $f$, for the austenite/martensite microstructure in $Li_2Mn_2O_4$. Supplementary Table 5 lists these solutions using $\mathbf{U}_1$ and $\mathbf{U}_2$ as the representative stretch tensors in Eq. (S.6).

| Volume fraction | Solutions of austenite-martensite interface | | |
|---|---|---|---|
| $f = 0.2158$ $k = 1$ | $\mathbf{b}_1 = [0.0996, 0.0043, 0.0646]$ $\mathbf{\hat{m}_1} = [0.8653, 0.0375, -0.4998]$ | $\mathbf{Q}'_1 =$ | $\begin{bmatrix} 0.9982 & -0.0316 & -0.0514 \\ 0.0315 & 0.9995 & -0.0022 \\ 0.0514 & 0.0006 & 0.9987 \end{bmatrix}$ |
| $f = 0.2158$ $k = -1$ | $\mathbf{b}_2 = [-0.0996, -0.0043, 0.0646]$ $\mathbf{\hat{m}}_2 = [-0.8653, -0.0375, -0.4998]$ | $\mathbf{Q}'_2 =$ | $\begin{bmatrix} 0.9982 & -0.0316 & 0.0514 \\ 0.0315 & 0.9995 & 0.0022 \\ -0.0514 & -0.0006 & 0.9987 \end{bmatrix}$ |
| $f = 0.7842$ $k = 1$ | $\mathbf{b}_3 = [0.0043, 0.0996, 0.0646]$ $\mathbf{\hat{m}}_3 = [0.0375, 0.8653, -0.4998]$ | $\mathbf{Q}'_3 =$ | $\begin{bmatrix} 0.9191 & -0.0288 & -0.0022 \\ 0.0339 & 1.0855 & -0.0514 \\ 0.0038 & 0.0558 & 0.9987 \end{bmatrix}$ |
| $f = 0.7842$ $k = -1$ | $\mathbf{b}_4 = [-0.0043, -0.0996, 0.0646]$ $\mathbf{\hat{m}}_4 = [-0.0375, -0.8653, -0.4998]$ | $\mathbf{Q}'_4 =$ | $\begin{bmatrix} 0.9191 & -0.0288 & 0.0022 \\ 0.0339 & 1.0855 & 0.0514 \\ -0.0038 & -0.0558 & 0.9987 \end{bmatrix}$ |

Supplementary Table 5: Four solutions for $Li_2Mn_2O_4$ austenite-martensite interface with $\mathbf{U}_1$ and $\mathbf{U}_2$ variants

### Self-accommodating microstructure

Self-accommodation is a material's ability to undergo phase transformations without any macroscopic change in its volume, despite crystal structure changes at the atomic scale, see Supplementary Figure 4(b). As noted previously, phase transformations are accompanied by an abrupt change in lattice geometries that induces coherency or misfit strains between neighboring phases. However, in some phase transformation materials (e.g., shape memory alloys), lattice variants can arrange themselves to form a compatible and stress-free microstructure that does not induce coherency stresses at the phase boundary. These microstructures accommodate the original shape of the material and thus have zero macroscopic volume changes. These self-accommodating microstructures, if stabilized in intercalation cathodes, would address the structural degradation problems related to volume changes and interfacial stresses in active particles.

Self-accommodating microstructures form in phase transformation materials that satisfy very specific lattice geometry conditions. Following Bhattacharya [1], the necessary and sufficient condi-

tions to generate self-accommodating microstructures are of two types: (a) There must be volume preservation during phase transformation, i.e., $\det \mathbf{U} = 1$; (b) For non-cubic symmetries of the reference phase, there must be no stretch along the $c-$axis of the lattice during transformation.

Supplementary Table. 6 lists the necessary and sufficient conditions to form self-accommodation microstructures. For cubic symmetry, $\det \mathbf{U} = 1$ is the only necessary and sufficient condition. For tetragonal, orthorhombic and hexagonal symmetry in the reference phase, other than satisfying $\det \mathbf{U} = 1$, more conditions are constrained on elements $D_{ij}$ of the matrix $\mathbf{D} = \mathbf{U}^{\mathrm{T}} \mathbf{U}$. For monoclinic symmetry, we use $\det \mathbf{M} = 1$ as the necessary and sufficient condition, and further details can be found in Ref. [1].

| Symmetry of the reference phase | Necessary and sufficient conditions |
|---|---|
| Cubic | $\det \mathbf{U} = 1$ |
| Tetragonal | (1) $\det \mathbf{U} = 1$ <br> (2) $\frac{1}{D_{11}D_{22}-D_{12}^2} \leq 1 \leq D_{33}$ |
| Orthorhombic | (1) $\det \mathbf{U} = 1$ <br> (2) $\frac{1}{D_{11}D_{22}-D_{12}^2} \leq 1 \leq D_{33}$ <br> (3) $\frac{1}{D_{22}D_{33}-D_{23}^2} \leq 1 \leq D_{11}$ <br> (4) $\frac{1}{D_{11}D_{33}-D_{13}^2} \leq 1 \leq D_{22}$ |
| Monoclinic | $\det \mathbf{M} = 1$ <br> For further details, see Ref. [1] Section 3.4 |
| Hexagonal | (1) $\det \mathbf{U} = 1$ <br> (2) $\frac{1}{D_{11}D_{22}-D_{12}^2} \leq 1 \leq D_{33}$ |

Supplementary Table 6: Necessary and sufficient conditions for forming self-accommodating microstructures. $\mathbf{U}$ is the transformation stretch matrix. $\mathbf{D} = \mathbf{U}^{\mathrm{T}} \mathbf{U}$ and $D_{ij}$ are the components of $\mathbf{D}$. In monoclinic symmetry, $\det \mathbf{M} = 1$ is used as the necessary and sufficient ccondition.

## $\lambda_2 = 1$ microstructure

A special case of Eq. (S.6) is at volume fractions $f = 0$ and $f = 1$. With these volume fractions, a planar interface forms between the reference phase and a *single* variant of the intercalated phase.

This phase boundary is exactly compatible and would have zero interfacial stresses. These stress-free phase boundaries, if stabilized in intercalation materials, can move back and forth during (dis)charge processes without inducing significant internal stresses.

These stress-free microstructures form when the reference and intercalated phases satisfy the kinematic compatibility condition. From Eq. (S.6) with $f \in \{0, 1\}$ and suitably redefining $\mathbf{Q}$ and $\mathbf{U}$, we have:

$$\mathbf{QU} - \mathbf{I} = \mathbf{b} \otimes \hat{\mathbf{m}} \qquad (S.7)$$

Eq. (S.7) has solution if and only if the middle eigenvalue $\lambda_2 = 1$, and the other two eigenvalues satisfy $0 < \lambda_1 \leq 1$ and $\lambda_3 \geq 1$. The solution to Eq. (S.7) describes a coherent plane between the reference phase and a single variant of the transformed phase.[5] These stress-free $\lambda_2 = 1$ microstructures have been shown to dramatically lower fatigue. For example, in Ti-Ni-Cu shape-memory alloys, another material, the $\lambda_2 = 1$ microstructures have enhanced its reversible phase transformation cycles for over ten million times with ultra-low fatigue [7].

Besides the $\lambda_2 = 1$ condition, researchers have identified other compatibility conditions, referred to as the cofactor conditions, which permit stress-free microstructures to form during phase transformations. Materials satisfying these cofactor conditions have solutions to Eq. (S.6) for any volume fraction $0 \leq f \leq 1$, and the resulting microstructures are stress-free [8]. Among the cofactor conditions listed in Ref. [8], the $\lambda_2 = 1$ condition has emerged as a critical condition for enhanced reversibility.

Eqs. (S.3), (S.6) and (S.7) serve as the design principles necessary to form shape-memory-like microstructures in crystalline materials. These design principles establish a direct link between individual lattice deformations and continuum microstructures in phase transformation materials. These principles can be used to systematically investigate whether any known intercalation material can form shape-memory-like microstructures.

---

[5]Note, this differs from the twin interface in Eq. (S.3) that connects two variants of the *same* phase.

## Additional results for Study 1

In this section, we summarize the stretch tensors and geometric constraints of commonly used intercalation cathodes $n = 25$ that we analyzed in Study 1 of the main paper. These results are also shown in Figs. 5-7 of the main paper, and the values are listed below as additional information.

Supplementary Table 7 lists the stretch tensors of $n = 25$ intercalation cathodes computed using our theoretical framework in Study 1. These stretch tensors are computed using structural data input from XRD experiments in Refs. [3, 9–30]. For each compound, we identify its symmetry lowering transformations that are summarized in Fig. 5 of the main paper.

| Family | Compound | Exp. Bravais lattice[#] | Ref. | Predicted symmetry change | Predicted stretch tensor |
|---|---|---|---|---|---|
| Layered | $CoO_2$ | $hP$ | [9] | Hexagonal $\rightarrow$ | $\begin{bmatrix} 0.9996 & 0 & 0 \\ 0 & 0.9996 & 0 \\ 0 & 0 & 1.0970 \end{bmatrix}$ |
| | $LiCoO_2$ | $hR$ | [9] | Hexagonal | |
| | $NiO_2$ | $hR$ | [10] | Hexagonal $\rightarrow$ | $\begin{bmatrix} 1.0249 & 0 & 0 \\ 0 & 1.0249 & 0 \\ 0 & 0 & 1.0527 \end{bmatrix}$ |
| | $LiNiO_2$ | $hR$ | [10] | Hexagonal | |
| Spinel | $LiMn_2O_4$ | $cF$ | [11] | *Cubic $\rightarrow$ | $\begin{bmatrix} 0.9690 & 0 & 0 \\ 0 & 0.9690 & 0 \\ 0 & 0 & 1.1226 \end{bmatrix}$ |
| | $Li_2Mn_2O_4$ | $tI$ | [11] | Tetragonal | |
| | $LiCrMnO_4$ | $cF$ | [12] | *Cubic $\rightarrow$ | $\begin{bmatrix} 0.9958 & 0 & 0 \\ 0 & 0.9958 & 0 \\ 0 & 0 & 1.0583 \end{bmatrix}$ |
| | $Li_2CrMnO_4$ | $tI$ | [12] | Tetragonal | |
| | $Ni_{0.5}Mn_{1.5}O_4$ | $cF$ | [13] | Cubic $\rightarrow$ | $\begin{bmatrix} 0.9695 & 0 & 0 \\ 0 & 0.9695 & 0 \\ 0 & 0 & 0.9695 \end{bmatrix}$ |
| | $LiNi_{0.5}Mn_{1.5}O_4$ | $cF$ | [13] | Cubic | |
| Olivine | $FePO_4$ | $oP$ | [14] | Orthorhombic $\rightarrow$ | $\begin{bmatrix} 1.0593 & 0 & 0 \\ 0 & 1.0730 & 0 \\ 0 & 0 & 1.0336 \end{bmatrix}$ |
| | $NaFePO_4$ | $oP$ | [14] | Orthorhombic | |
| | $MnPO_4$ | $oP$ | [15] | Orthorhombic $\rightarrow$ | $\begin{bmatrix} 1.0971 & 0 & 0 \\ 0 & 1.0735 & 0 \\ 0 & 0 & 1.0473 \end{bmatrix}$ |
| | $NaMnPO_4$ | $oP$ | [16] | Orthorhombic | |

| Olivine | MnPO$_4$ | $oP$ | [15] | Orthorhombic $\rightarrow$ | $\begin{bmatrix} 1.0857 & 0 & 0 \\ 0 & 1.0343 & 0 \\ 0 & 0 & 0.9945 \end{bmatrix}$ |
| --- | --- | --- | --- | --- | --- |
|  | LiMnPO$_4$ | $oP$ | [15] | Orthorhombic |  |
|  | FePO$_4$ | $oP$ | [17] | Orthorhombic $\rightarrow$ | $\begin{bmatrix} 1.0373 & 0 & 0 \\ 0 & 1.0522 & 0 \\ 0 & 0 & 0.9802 \end{bmatrix}$ |
|  | LiFePO$_4$ | $oP$ | [17] | Orthorhombic |  |
|  | CoPO$_4$ | $oP$ | [18] | Orthorhombic $\rightarrow$ | $\begin{bmatrix} 1.0112 & 0 & 0 \\ 0 & 1.0118 & 0 \\ 0 & 0 & 0.9958 \end{bmatrix}$ |
|  | LiCoPO$_4$ | $oP$ | [18] | Orthorhombic |  |
| Tavorite | $\epsilon$-VOPO$_4$ | $mP$ | [19] | Monoclinic $\rightarrow$ | $\begin{bmatrix} 0.9911 & 0 & -0.0025 \\ 0 & 1.1440 & 0.0142 \\ -0.0025 & 0.0142 & 0.9238 \end{bmatrix}$ |
|  | $\alpha$-LiVOPO$_4$ | $aP$ | [19] | Triclinic |  |
|  | FeSO$_4$F | $aP$ | [20] | Triclinic $\rightarrow$ | $\begin{bmatrix} 1.0184 & 0.0555 & -0.0042 \\ 0.0555 & 1.0760 & 0.0286 \\ -0.0042 & 0.0286 & 1.0217 \end{bmatrix}$ |
|  | LiFeSO$_4$F | $aP$ | [20] | Triclinic |  |
|  | FeSO$_4$F | $mC$ | [21] | Triclinic $\rightarrow$ | $\begin{bmatrix} 1.1910 & 0 & -0.0023 \\ 0 & 0.9426 & 0.0351 \\ -0.0023 & 0.0351 & 1.0402 \end{bmatrix}$ |
|  | NaFeSO$_4$F | $mP$ | [21] | Monoclinic |  |
| NASICON | Fe$_2$(MoO$_4$)$_3$ | $mP$ | [3] | Monoclinic $\rightarrow$ | $\begin{bmatrix} 1.0116 & 0 & 0.0080 \\ 0 & 1.0269 & 0 \\ 0.0080 & 0 & 1.0196 \end{bmatrix}$ |
|  | Li$_2$Fe$_2$(MoO$_4$)$_3$ | $oP$ | [3] | Monoclinic |  |

| | | | | | | | | |
|---|---|---|---|---|---|---|---|---|
| NASICON | Fe$_2$(WO$_4$)$_3$ | $mP$ | [22] | Monoclinic $\to$ | $\begin{bmatrix} 0.9962 & 0 & 0.0085 \\ 0 & 1.0233 & 0 \\ 0.0085 & 0 & 1.0234 \end{bmatrix}$ | | | |
| | Li$_2$Fe$_2$(WO$_4$)$_3$ | $oP$ | [22] | Monoclinic | | | | |
| Perovskite | ReO$_3$ | $tP$ | [23] | Tetragonal $\to$ | $\begin{bmatrix} 0.9645 & 0 & 0 \\ 0 & 0.9645 & 0 \\ 0 & 0 & 1.0281 \end{bmatrix}$ | | | |
| | LiReO$_3$ | $tP$ | [23] | Tetragonal | | | | |
| | LiReO$_3$ | $tP$ | [23] | Tetragonal $\to$ | $\begin{bmatrix} 0.9744 & 0 & 0 \\ 0 & 0.9744 & 0 \\ 0 & 0 & 1.0956 \end{bmatrix}$ | | | |
| | Li$_2$ReO$_3$ | $tP$ | [23] | Tetragonal | | | | |
| Pyrophosphate | LiFeP$_2$O$_7$ | $mP$ | [24] | Monoclinic $\to$ | $\begin{bmatrix} 0.9667 & 0 & 0.0033 \\ 0 & 1.0157 & 0 \\ 0.0033 & 0 & 0.9991 \end{bmatrix}$ | | | |
| | Li$_2$FeP$_2$O$_7$ | $mP$ | [24] | Monoclinic | | | | |
| | VP$_2$O$_7$ | $mP$ | [25] | Monoclinic $\to$ | $\begin{bmatrix} 1.0072 & 0 & -0.0254 \\ 0 & 1.0310 & 0 \\ -0.0254 & 0 & 0.9964 \end{bmatrix}$ | | | |
| | LiVP$_2$O$_7$ | $mP$ | [25] | Monoclinic | | | | |
| | NaFe$_3$(PO$_4$)$_2$(P$_2$O$_7$) | $oP$ | [26] | Orthorhombic $\to$ | $\begin{bmatrix} 1.0231 & 0 & 0 \\ 0 & 1.0209 & 0 \\ 0 & 0 & 0.9951 \end{bmatrix}$ | | | |
| | Na$_4$Fe$_3$(PO$_4$)$_2$(P$_2$O$_7$) | $oP$ | [26] | Orthorhombic | | | | |
| | $\beta$-NaFeP$_2$O$_7$ | $aP$ | [27] | Triclinic $\to$ | $\begin{bmatrix} 1.0185 & -0.0108 & -0.0254 \\ -0.0108 & 1.0068 & -0.0156 \\ -0.0254 & -0.0156 & 1.0091 \end{bmatrix}$ | | | |
| | Na$_2$FeP$_2$O$_7$ | $aP$ | [27] | Triclinic | | | | |

| | | | | | |
|---|---|---|---|---|---|
| Fluorophosphate | VPO$_4$F | $mC$ | [28] | Monoclinic $\to$ | $\begin{bmatrix} 1.0607 & 0.0090 & 0.0155 \\ 0.0090 & 0.9648 & 0.0107 \\ 0.0155 & 0.0107 & 1.0683 \end{bmatrix}$ |
| | LiVPO$_4$F | $aP$ | [28] | Triclinic | |
| | LiVPO$_4$F | $aP$ | [28] | *Triclinic $\to$ | $\begin{bmatrix} 1.0110 & 0.0181 & -0.0186 \\ 0.0181 & 1.0360 & -0.0259 \\ -0.0186 & -0.0259 & 1.0270 \end{bmatrix}$ |
| | Li$_2$VPO$_4$F | $mC$ | [28] | Monoclinic | |
| | VPO$_4$F | $mC$ | [29] | Monoclinic $\to$ | $\begin{bmatrix} 1.0849 & -0.0105 & 0.0356 \\ -0.0105 & 0.9243 & 0.0063 \\ 0.0356 & 0.0063 & 1.1305 \end{bmatrix}$ |
| | NaVPO$_4$F | $mC$ | [29] | Triclinic | |
| | NaFePO$_4$F | $oP$ | [30] | Orthorhombic $\to$ | $\begin{bmatrix} 1.0217 & 0 & 0 \\ 0 & 0.9808 & 0 \\ 0 & 0 & 1.0351 \end{bmatrix}$ |
| | Na$_2$FePO$_4$F | $oP$ | [30] | Orthorhombic | |

Supplementary Table 7: Database of representative intercalation compounds. # The symbols for Bravais lattices follow the nomenclature used in the International Union of Crystallography tables [31]. This column lists the experimentally documented Bravais lattice symmetries from Ref. [3, 9–30]. Specifically, $c$, $h$, $t$, $o$, $m$ and $a$ stand for cubic, hexagonal, tetragonal, orthorhombic, monoclinic and triclinic respectively. *Our theoretical framework predicts multiple structural transformation pathways for these compounds. In our calculations, we choose the structural transformation pathway (i.e., the stretch tensor) that describes a symmetry lowering transformation. This is because these transformations generate two or more variants and are suitable to form energy minimizing twin deformation microstructures.

Supplementary Table. 8 lists the middle eigenvalues and total volume of intercalation cathodes that undergo a symmetry lowering transformation in Study 1. These results are summarized in the polar plots shown in Fig. 5 of the main paper.

| Transformation | Symmetry | Eigenvalues of $\mathbf{C}$ | $\lambda_2(\mathbf{I}/\mathbf{U})$ | $\det \mathbf{U}$ |
|---|---|---|---|---|
| $\text{Li}_x\text{Mn}_2\text{O}_4$ $x = 1 \to 2$ | Cubic $\to$ Tetragonal $N = 3$ | $\{1.3421, 1, 0.7451\}$ | 0.9390 | 1.0541 |
| $\text{Li}_x\text{CrMnO}_4$ $x = 1 \to 2$ | Cubic $\to$ Tetragonal $N = 3$ | $\{1.1296, 1, 0.8853\}$ | 0.9916 | 1.0494 |
| $\alpha\text{-Li}_x\text{VOPO}_4$ $x = 0 \to 1$ | Monoclinic $\to$ Triclinic $N = 2$ | $\{1.0571, 1, 0.9459\}$ | 0.9824 | 1.0472 |
| $\text{Li}_x\text{VPO}_4\text{F}$ $x = 0 \to 1$ | Monoclinic $\to$ Triclinic $N = 2$ | $\{1.0564, 1, 0.9466\}$ | 1.0996 | 1.0928 |
| $\text{Na}_x\text{VPO}_4\text{F}$ $x = 0 \to 1$ | Monoclinic $\to$ Triclinic $N = 2$ | $\{1.0508, 1, 0.9516\}$ | 1.1374 | 1.1323 |
| $\text{Na}_x\text{FeSO}_4\text{F}$ $x = 1 \to 0$ | Monoclinic $\to$ Triclinic $N = 2$ | $\{1.0072, 1, 0.9928\}$ | 0.9045 | 0.8574 |
| $\text{Li}_x\text{VPO}_4\text{F}$ $x = 2 \to 1$ | Monoclinic $\to$ Triclinic $N = 2$ | $\{1.0624, 1, 0.9413\}$ | 0.9880 | 0.9308 |

Supplementary Table 8: Candidate cathode materials from the database of representative intercalation compounds. Here eigenvalues of $\mathbf{C} = (\mathbf{FG}^{-1})^{\text{T}}\mathbf{FG}^{-1}$ can be derived by solving Eq. (S.3). $\lambda_2$ value can be obtained by solving Eq. (S.7).

Supplementary Table. 9 lists the self-accommodation geometric constraints of cathode materials, which have a monoclinic symmetry in the reference phase. We note that although some intercalation cathodes satisfy $|\det \mathbf{M} - 1| \to 0$, these compounds do not satisfy the other geometric constraints $0 < M_{11} < 1 + \delta^2$ and $0 < M_{33} < 1$ necessary to form self-accommodating microstructures.

| Transformation | $\mathbf{M}$ | $\det \mathbf{M}$ | $1 + \delta^2$ | $M_{11}$ | $M_{33}$ |
|---|---|---|---|---|---|
| $\alpha$-Li$_{0\to 1}$VOPO$_4$ | $\begin{bmatrix} 1.0181 & -0.0003 & 0.0056 \\ -0.0003 & 0.7676 & -0.0589 \\ 0.0056 & -0.0589 & 1.1715 \end{bmatrix}$ | 0.9119 | 1.0001 | 1.0181 | 1.1715 |
| Li$_{0\to 1}$VPO$_4$F | $\begin{bmatrix} 0.8908 & -0.0361 & -0.0239 \\ -0.0361 & 1.0741 & -0.0425 \\ -0.0239 & -0.0425 & 0.8787 \end{bmatrix}$ | 0.8373 | 1.0008 | 0.8908 | 0.8787 |
| Na$_{0\to 1}$VPO$_4$F | $\begin{bmatrix} 0.8544 & 0.0465 & -0.0539 \\ 0.0465 & 1.1702 & -0.0279 \\ -0.0539 & -0.0279 & 0.7857 \end{bmatrix}$ | 0.7799 | 1.0006 | 0.8544 | 0.7857 |
| Na$_{1\to 0}$FeSO$_4$F | $\begin{bmatrix} 0.8952 & 0.0007 & 0.0766 \\ 0.0007 & 1.4186 & 0.0083 \\ 0.0766 & 0.0083 & 1.0777 \end{bmatrix}$ | 1.3603 | 1 | 0.8952 | 1.0777 |
| Li$_{2\to 1}$VPO$_4$F | $\begin{bmatrix} 1.1181 & -0.0568 & 0.0478 \\ -0.0568 & 1.0041 & -0.0266 \\ 0.0478 & -0.0266 & 1.0337 \end{bmatrix}$ | 1.1543 | 1.0009 | 1.1181 | 1.0337 |

Supplementary Table 9: Self-accommodating condition check for candidate cathode materials with a monoclinic reference phase. More details can be found in Ref. [1] Section 3.4.

Supplementary Table. 10 lists the cofactor conditions for intercalation cathodes (Study 1) that undergo symmetry lowering transformation. Of these conditions, $\lambda_2 = 1$ is found to be the most dominant condition [8] and the results are summarized in Fig. 5(b) of the main paper.

| Transformation | $\lambda_2$ | $\mathbf{a} \cdot \mathbf{U}\mathrm{cof}(\mathbf{U}^2 - \mathbf{I})\mathbf{n} = 0$ | $\mathrm{tr}\mathbf{U}^2 - \det\mathbf{U}^2 - \frac{|\mathbf{a}|^2|\mathbf{n}|^2}{4} - 2 \geq 0$ |
|---|---|---|---|
| $\mathrm{Li}_{1\to 2}\mathrm{Mn}_2\mathrm{O}_4$ | 0.9390 | 0.0031 | 0.0036 |
| $\mathrm{Li}_{1\to 2}\mathrm{CrMnO}_4$ | 0.9916 | 0.0001 | $-0.0020$ |
| $\alpha$-$\mathrm{Li}_{0\to 1}\mathrm{VOPO}_4$ | 0.9824 | 0 | 0.0472 |
| $\mathrm{Li}_{0\to 1}\mathrm{VPO}_4\mathrm{F}$ | 1.0996 | $-0.0002$ | 0.0029 |
| $\mathrm{Na}_{0\to 1}\mathrm{VPO}_4\mathrm{F}$ | 1.1374 | $-0.0004$ | 0.0294 |
| $\mathrm{Na}_{1\to 0}\mathrm{FeSO}_4\mathrm{F}$ | 0.9045 | 0 | 0.0273 |
| $\mathrm{Li}_{2\to 1}\mathrm{VPO}_4\mathrm{F}$ | 0.9880 | 0 | $-0.0020$ |

Supplementary Table 10: Cofactor condition check for candidate cathode materials. More information can be found in [8].

| Compounds | Voltage window | $\|\det \mathbf{U} - 1\|/$ $\|\det \mathbf{M} - 1\|$ | $\|\lambda_2 - 1\|$ | # of cycles | C rate | Capacity retention | Ref. | Notes |
|---|---|---|---|---|---|---|---|---|
| $\beta - \text{Li}_{1-2}\text{IrO}_3$ | 2.0V-4.0V | 0.0045 | 0.0190 | 30 | C/10 | 95 % | [32] | |
| $\beta - \text{Li}_{0.35-1}\text{IrO}_3$ | 4.0V-4.45V | 0.0132 | 0.0170 | 30 | C/10 | 86 % | [32] | |
| $\text{Na}_{1-2}\text{FeP}_2\text{O}_7$ | 2.5V-4.0V | 0.0337 | 0.0372 | 1000 | 100 mA/g | 91% | [33] | Solid solution |
| $\text{Li}_{0.4-2}\text{Ru}_{0.75}\text{Sn}_{0.25}\text{O}_3$ (LRSO) | 2.0V-4.6V | 0.0340 | 0.0311 | 60 | C/10 | 88% | [34] | Solid solution |
| $\text{Li}_{0.14-1}\text{Ni}_{0.13}\text{Mn}_{0.54}\text{Co}_{0.13}\text{O}_2$ (Li-rich NMC-I) | 2.0V-4.8V | 0.0302 | 0.0202 | 25 | C/5 | 82 % | [35] | Solid solution |
| $\text{Na}_{1-4}\text{Fe}_3(\text{PO}_4)_2(\text{P}_2\text{O}_7)$ | 1.7V-4.5V | 0.0394 | 0.0423 | 100 | C/5 | 86% | [36] | |
| $\beta - \text{Li}_{0-0.35}\text{IrO}_3$ | 4.45V-4.8V | 0.0431 | 0.0590 | 30 | C/10 | 49 % | [32] | Electrolyte decomposition or poor reversibility |
| $\text{Na}_{0.5-1}\text{NbO}_2$ | 0.0V-2.0V | 0.0457 | 0.0068 | 600 | 30 mA/g | 82% | [37] | |
| $\text{Li}_{0.14-1.44}\text{Ni}_{0.13}\text{Mn}_{0.54}\text{Co}_{0.13}\text{O}_2$ (Li-rich NMC-II) | 1.2V-4.8V | 0.0517 | 0.0390 | 25 | C/5 | 68 % | [35] | Irreversibile transformation $P' \rightarrow P''$ around 1.4 V |
| $\text{Li}_{1-2}\text{Mn}_2\text{O}_4$ | 2.2V-3.3V | 0.0541 | 0.0610 | 10 | 0.5mA/cm$^2$ | 42 % | [38] | Irreversible transformation [5] |
| $\text{Li}_{1-3}\text{Ti}_2(\text{PO}_4)_3$ | 2.0V-3.4V | 0.0651 | 0.0280 | 40 | C/10 | 77% | [39] | |
| $\text{Li}_{0.2-1.2}\text{Ti}_{0.4}\text{Mn}_{0.4}\text{O}_2$ (LTMO) | 1.5V-4.8V | 0.0653 | 0.0431 | 50 | 10 mA/g | 65% | [40] | |
| $\text{Li}_{0-1}\text{FePO}_4$ | 2.0V-4.5V | 0.0699 | 0.0760 | 100 | 750 mA/g | 64 % | [41] | |
| $\text{Li}_{0-1}\text{Mn}_2\text{O}_4$ | 3.5V-4.3V | 0.0846 | 0.0556 | 200 | C/2 | 61 % | [42] | Sol-gel Li$_x$Mn$_2$O$_4$ |
| $\text{Li}_{0-1}\text{FeSO}_4\text{F}$ | 2.5V-4.2V | 0.1156 | 0.0496 | 50 | C/10 | 92% | [20] | Ionothermal synthesis |
| $\text{Na}_{0-1}\text{FePO}_4$ | 2.2V-4.3V | 0.1749 | 0.1221 | 50 | 7.5 mA/g | 88 % | [43] | Amorphous phase exists [44] |
| $\text{Na}_{0-1}\text{VPO}_4\text{F}$ | 3.0V-4.5V | 0.2201 | 0.1374 | 20 | 10 mA/g | 74 % | [45] | |
| $\text{Na}_{0-1}\text{MnPO}_4$ | 2.5V-4.5V | 0.2334 | 0.1525 | 20 | C/20 | 64% | [46] | Covered with carbon black |

Supplementary Table 11: Performance summary of intercalation compounds under cycling.

# References

1. Bhattacharya, K. Self-accommodation in martensite. *Archive for Rational Mechanics and Analysis* **120,** 201–244 (1992).

2. Chen, X., Song, Y., Tamura, N. & James, R. D. Determination of the stretch tensor for structural transformations. *Journal of the Mechanics and Physics of Solids* **93,** 34–43 (2016).

3. Zhou, S., Barim, G., Morgan, B. J., Melot, B. C. & Brutchey, R. L. Influence of rotational distortions on $Li^+$-and $Na^+$-intercalation in anti-NASICON $Fe_2(MoO_4)_3$. *Chemistry of Materials* **28,** 4492–4500 (2016).

4. Bhattacharya, K. *et al. Microstructure of martensite: why it forms and how it gives rise to the shape-memory effect* (Oxford University Press, 2003).

5. Erichsen, T., Pfeiffer, B., Roddatis, V. & Volkert, C. A. Tracking the diffusion-controlled lithiation reaction of $LiMn_2O_4$ by In Situ TEM. *ACS Applied Energy Materials* **3,** 5405–5414 (2020).

6. Ball, J. M. & James, R. D. in *Analysis and Continuum Mechanics* 647–686 (Springer, 1989).

7. Chluba, C. *et al.* Ultralow-fatigue shape memory alloy films. *Science* **348,** 1004–1007 (2015).

8. Chen, X., Srivastava, V., Dabade, V. & James, R. D. Study of the cofactor conditions: conditions of supercompatibility between phases. *Journal of the Mechanics and Physics of Solids* **61,** 2566–2587 (2013).

9. Tarascon, J. *et al.* In situ structural and electrochemical study of $Ni_{1-x}Co_xO_2$ metastable oxides prepared by soft chemistry. *Journal of Solid State Chemistry* **147,** 410–420 (1999).

10. Ohzuku, T., Ueda, A. & Nagayama, M. Electrochemistry and structural chemistry of $LiNiO_2$(R3m) for 4 volt secondary lithium cells. *Journal of the Electrochemical Society* **140,** 1862 (1993).

11. Mukhopadhyay, A. & Sheldon, B. W. Deformation and stress in electrode materials for Li-ion batteries. *Progress in Materials Science* **63,** 58–116 (2014).

12. Kano, S. & Sato, M. Structure and lithium insertion characteristics of $LiCrMnO_4$. *Solid State Ionics* **79,** 215–219 (1995).

13. Choi, H. W., Kim, S. J., Rim, Y.-H. & Yang, Y. S. Effect of Lithium Deficiency on Lithium-Ion Battery Cathode $LiNi_{0.5}Mn_{1.5}O_4$. *The Journal of Physical Chemistry C* **119,** 27192–27199 (2015).

14. Moreau, P., Guyomard, D., Gaubicher, J. & Boucher, F. Structure and stability of sodium intercalated phases in olivine $FePO_4$. *Chemistry of Materials* **22,** 4126–4128 (2010).

15. Mishima, Y. *et al.* MEM Charge Density Study of Olivine $LiMPO_4$ and $MPO_4$ (M = Mn, Fe) as Cathode Materials for Lithium-Ion Batteries. *The Journal of Physical Chemistry C* **117,** 2608–2615 (2013).

16. Lee, K. T., Ramesh, T., Nan, F., Botton, G. & Nazar, L. F. Topochemical synthesis of sodium metal phosphate olivines for sodium-ion batteries. *Chemistry of Materials* **23,** 3593–3600 (2011).

17. Padhi, A. K., Nanjundaswamy, K. S. & Goodenough, J. B. Phospho-olivines as positive-electrode materials for rechargeable lithium batteries. *Journal of the electrochemical society* **144,** 1188 (1997).

18. Amine, K., Yasuda, H. & Yamachi, M. Olivine $LiCoPO_4$ as 4.8 V electrode material for lithium batteries. *Electrochemical and Solid-State Letters* **3,** 178 (2000).

19. Kerr, T., Gaubicher, J. & Nazar, L. Highly Reversible Li Insertion at 4 V in $\varepsilon$-$VOPO_4$/$\alpha$-$LiVOPO_4$ Cathodes. *Electrochemical and Solid-State Letters* **3,** 460 (2000).

20. Recham, N. *et al.* A 3.6 V lithium-based fluorosulphate insertion positive electrode for lithium-ion batteries. *Nature materials* **9,** 68–74 (2010).

21. Tripathi, R., Ramesh, T., Ellis, B. L. & Nazar, L. F. Scalable synthesis of tavorite $LiFeSO_4F$ and $NaFeSO_4F$ cathode materials. *Angewandte Chemie* **122,** 8920–8924 (2010).

22. Barim, G., Cottingham, P., Zhou, S., Melot, B. C. & Brutchey, R. L. Investigating the mechanism of reversible lithium insertion into anti-nasicon $Fe_2(WO_4)_3$. *ACS Applied Materials & Interfaces* **9,** 10813–10819 (2017).

23. Bashian, N. H. *et al.* Correlated polyhedral rotations in the absence of polarons during electrochemical insertion of lithium in $ReO_3$. *ACS Energy Letters* **3,** 2513–2519 (2018).

24. Nishimura, S.-i., Nakamura, M., Natsui, R. & Yamada, A. New lithium iron pyrophosphate as 3.5V class cathode material for lithium ion battery. *Journal of the American Chemical Society* **132,** 13596–13597 (2010).

25. Rousse, G. *et al.* Crystal structure of a new vanadium (IV) diphosphate: $VP_2O_7$, prepared by lithium extraction from $LiVP_2O_7$. *International Journal of Inorganic Materials* **3,** 881–887 (2001).

26. Kim, H. *et al.* New iron-based mixed-polyanion cathodes for lithium and sodium rechargeable batteries: combined first principles calculations and experimental study. *Journal of the American Chemical Society* **134,** 10369–10372 (2012).

27. Barpanda, P. *et al.* $Na_2FeP_2O_7$: a safe cathode for rechargeable sodium-ion batteries. *Chemistry of Materials* **25,** 3480–3487 (2013).

28. Ellis, B. L., Ramesh, T., Davis, L. J., Goward, G. R. & Nazar, L. F. Structure and electrochemistry of two-electron redox couples in lithium metal fluorophosphates based on the tavorite structure. *Chemistry of Materials* **23,** 5138–5148 (2011).

29. Boivin, E. *et al.* Vanadyl-type defects in Tavorite-like $NaVPO_4F$: from the average long range structure to local environments. *Journal of Materials Chemistry A* **5,** 25044–25055 (2017).

30. Lee, I. K., Shim, I.-B. & Kim, C. S. Phase transition studies of sodium deintercalated $Na_{2-x}FePO_4F$ ($0 \leq x \leq 1$) by Mössbauer spectroscopy. *Journal of Applied Physics* **109,** 07E136 (2011).

31. De Wolff, P. o. *et al.* Nomenclature for crystal families, Bravais-lattice types and arithmetic classes. Report of the International Union of Crystallography Ad-Hoc Committee on the Nomenclature of Symmetry. *Acta Crystallographica Section A: Foundations of Crystallography* **41,** 278–280 (1985).

32. Pearce, P. E. *et al.* Evidence for anionic redox activity in a tridimensional-ordered Li-rich positive electrode $\beta$-$Li_2IrO_3$. *Nature materials* **16,** 580–586 (2017).

33. Chen, C.-Y. *et al.* Pyrophosphate $Na_2FeP_2O_7$ as a low-cost and high-performance positive electrode material for sodium secondary batteries utilizing an inorganic ionic liquid. *Journal of Power Sources* **246,** 783–787 (2014).

34. Sathiya, M. *et al.* Reversible anionic redox chemistry in high-capacity layered-oxide electrodes. *Nature materials* **12,** 827–835 (2013).

35. Yin, W. *et al.* Structural evolution at the oxidative and reductive limits in the first electrochemical cycle of $Li_{1.2}Ni_{0.13}Mn_{0.54}Co_{0.13}O_2$. *Nature Communications* **11,** 1–11 (2020).

36. Kim, H. *et al.* Anomalous Jahn–Teller behavior in a manganese-based mixed-phosphate cathode for sodium ion batteries. *Energy & Environmental Science* **8,** 3325–3335 (2015).

37. Wang, X. *et al.* Anti-P2 structured $Na_{0.5}NbO_2$ and its negative strain effect. *Energy & Environmental Science* **8,** 2753–2759 (2015).

38. Thackeray, M. M. Manganese oxides for lithium batteries. *Progress in Solid State Chemistry* **25,** 1–71 (1997).

39. Patoux, S. & Masquelier, C. Lithium insertion into titanium phosphates, silicates, and sulfates. *Chemistry of Materials* **14,** 5057–5068 (2002).

40. Chen, D. *et al.* Role of fluorine in chemomechanics of cation-disordered rocksalt cathodes. *Chemistry of Materials* **33,** 7028–7038 (2021).

41. Wang, D., Li, H., Shi, S., Huang, X. & Chen, L. Improving the rate performance of $LiFePO_4$ by Fe-site doping. *Electrochimica Acta* **50,** 2955–2958 (2005).

42. Shaju, K. M. & Bruce, P. G. A stoichiometric nano-$LiMn_2O_4$ spinel electrode exhibiting high power and stable cycling. *Chemistry of Materials* **20,** 5557–5562 (2008).

43. Oh, S.-M., Myung, S.-T., Hassoun, J., Scrosati, B. & Sun, Y.-K. Reversible $NaFePO_4$ electrode for sodium secondary batteries. *Electrochemistry communications* **22,** 149–152 (2012).

44. Xiang, K. *et al.* Accommodating high transformation strains in battery electrodes via the formation of nanoscale intermediate phases: operando investigation of olivine $NaFePO_4$. *Nano letters* **17,** 1696–1702 (2017).

45. Zhuo, H. *et al.* The preparation of NaV$_{1-x}$Cr$_x$PO$_4$F cathode materials for sodium-ion battery. *Journal of power sources* **160,** 698–703 (2006).

46. Boyadzhieva, T. *et al.* Competitive lithium and sodium intercalation into sodium manganese phospho-olivine NaMnPO$_4$ covered with carbon black. *RSC advances* **5,** 87694–87705 (2015).


\end{document}